\documentclass[10pt,journal,compsoc]{IEEEtran}
\usepackage[ruled,lined,linesnumbered]{algorithm2e}
    \SetAlFnt{\small}
    \SetAlCapFnt{\small}
    \SetAlCapNameFnt{\small}
    \SetAlCapHSkip{0pt}
    \IncMargin{-\parindent}
\usepackage{amsmath}
\usepackage{amssymb}
\usepackage[english]{babel}
\usepackage{balance}
    \usepackage{silence}
\usepackage{caption}
\ifCLASSOPTIONcompsoc
\usepackage[nocompress]{cite}
\else
\usepackage{cite}
\fi
\usepackage{cleveref}
    \crefname{lstlisting}{listing}{listings}
    \Crefname{lstlisting}{Listing}{Listings}
\usepackage{color}
\usepackage{enumitem}
    \newlist{myenumi}{description}{10}
    \setlist[myenumi]{labelindent=\parindent, leftmargin=*, label=(\roman*), align=left}
    \setlist[myenumi]{leftmargin=0pt}
\usepackage{fancybox}
\usepackage{graphicx}
\usepackage{listings}
    \definecolor{dkgreen}{rgb}{0,0.6,0}
    \definecolor{gray}{rgb}{0.5,0.5,0.5}
    \definecolor{mauve}{rgb}{0.58,0,0.82}
\usepackage{multirow}
\usepackage{url}

\newcommand{\ea}{{et al.}}
\newcommand{\cf}{{\textit{cf.,}}}

\newcommand{\parrafo}[1]{\indent\textit{\textbf{#1:}}}

\newcommand{\eg}{\textit{e.g.,}}
\newcommand{\ie}{\textit{i.e.,}}
\newcommand{\hrs}{\emph{hybrid-RS}}
\newcommand{\duktape}{\emph{Duktape}}
\newcommand{\momit}{\emph{MoMIT}}

\newcommand{\photon}{\emph{Photon}}

\newcommand{\hypobox}[1]{\begin{center}%
\noindent\thicklines\setlength{\fboxsep}{8pt}%
\cornersize{0.2}\Ovalbox{\begin{minipage}{3.20in}%
\textit{#1}\end{minipage}} \end{center}}

\newcommand{\RQone}{\textbf{To what extent can \momit~miniaturize JS tests to run on constrained devices?} }
\newcommand{\RQtwo}{\textbf{What is the most convenient algorithm to instantiate \momit?}}

\newcommand{\PQone}{\textbf{Does the selected JS-interpreter features have an impact on software performance metrics?}}
\newcommand{\PQtwo}{\textbf{From the selected JS-interpreter features, which ones have the bigger impact on software performance metrics?}}

\begin{document}

\title{\momit~: Porting a JavaScript Interpreter\\on a Quarter Coin}

\author{Rodrigo~Morales,~\IEEEmembership{Member,~IEEE,}\\ Rub\'{e}n~Saborido,~\IEEEmembership{Member,~IEEE,}\\
and Yann-Ga{\"e}l Gu{\'e}h{\'e}neuc,~\IEEEmembership{Senior Member,~IEEE}
\IEEEcompsocitemizethanks{\IEEEcompsocthanksitem R.\ Morales, R.\ Saborido, Y.-G.\ Gu{\'e}h{\'e}neuc are with Concordia University Mont\'{e}al, Qu\'{e}bec, Canada. \protect\\
E-mail: rodrigomorales2@acm.org, rsain@uma.es, yann-gael.gueheneuc@concordia.ca.}

\thanks{Manuscript received March 2019.}}

\markboth{IEEE TRANSACTIONS ON SOFTWARE ENGINEERING,~Vol.~X, No.~X, January~2020}%
{Morales \MakeLowercase{\textit{et al.}}: \momit~}

\IEEEtitleabstractindextext{%
\begin{abstract}
The Internet of Things (IoT) is a network of physical, heterogeneous, connected devices providing services through private networks and the Internet. It connects a range of new devices to the Internet so they can communicate with Web servers and other devices around the world. Today's standard platform for communicating Web pages and Web apps is JavaScript (JS) and extending the same standard platform to connect IoT devices seems more than appropriate. However, porting JS applications to the large variety of IoT devices, specifically on System-on-a-Chip (SoCs) devices (\eg~Arduino Uno, Particle \photon), is challenging because these devices are constrained in terms of memory and storage capacity. Running JS applications adds an overhead of resources to deploy a code interpreter on the devices. Also, running JS applications may not be possible ``as is'' on some device missing some hardware/software capabilities. To address this problem, we propose \momit~a multiobjective optimization approach to miniaturize JS applications to run on IoT constrained devices. To validate \momit, we miniaturize a JS interpreter to execute a testbed comprised of 23 applications and measure their performances before and after applying the miniaturization process. We implement \momit~using three different search algorithms and found that it can reduce code size, memory usage, and CPU time by median values of 31\%, 56\%, and 36\% respectively.  Finally, \momit~ported the miniaturized JS interpreters up to to 2 SoCs additional devices, in comparison of using default JS interpreter features.
\end{abstract}

\begin{IEEEkeywords}
Internet of Things, Software Miniaturization, Multiobjective optimization, embedded devices, JavaScript, Evolutionary algorithms
\end{IEEEkeywords}}

\maketitle
\IEEEdisplaynontitleabstractindextext
\IEEEpeerreviewmaketitle

\IEEEraisesectionheading{\section{Introduction}
\label{sec:intro}}

\IEEEPARstart{T}{he Internet of Things} (IoT) is a network of physical, heterogeneous, connected devices providing services~\cite{ma2016networking} through private networks and the Internet. In 2016, Gartner\footnote{https://www.gartner.com/en/newsroom/press-releases/2016-01-14-gartner-says-by-2020-more-than-half-of-major-new-business-processes-and-systems-will-incorporate-some-element-of-the-internet-of-things} predicted that for 2018, 75\% of Internet of Things (IoT) projects will take up to twice as long as planned with insufficient staffing/expertise as the main cause of these delays. 

In the last years, product-centered companies realized the benefits of migrating to service-oriented systems whereby customers pay for a negotiated business outcome on a contract basis, instead of selling a single application. For example, Adobe and Microsoft are migrating their flagship products (\eg~Acrobat and Office) as Software-as-a-Service, transforming buyers into subscribers.  This impose new challenges as companies should be able to deploy their software in different architectures to reach as much customers as possible.
Due to the large variety of hardware and software architectures, from Cloud virtual machines to System-on-a-Chip (SoCs) devices, \eg{} the \photon, companies must spend lots of time and effort developing and maintaining their applications on different devices to reach their customers. Even if they decided to develop their applications using one programming language, they must remove some features to deploy them on the most constrained devices. 

Today's standard platform for communicating Web pages and Web apps is JavaScript (JS). A StackOverflow developers' survey positioned JS among the most popular programming languages for the last five years~\cite{sofdevsurvey2017}. On \textit{cloud architectures}, JS is commonly used in IoT to develop event-driven systems. It can handle large networks of connected devices and performs well when multiple tasks must be processed without waiting for others to complete (a typical scenario when developing websites). With major companies like IBM and Samsung using JS in their IoT projects, the demand for JS developers with IoT experience is high~\cite{itjobswatch, gooroo} and using the same JS platform to program IoT devices seems more than appropriate. However, IoT devices, particularly SoCs, possess limited memory and storage capacities to deploy a native JS interpreter, requiring the use of the C programming language to program them.

\parrafo{Objective} This paper proposes \momit, an automated multi-objective approach to \emph{miniaturize} existing JS applications to run on devices constrained in memory storage, and CPU capacity. Miniaturization is the process of removing unnecessary code features from applications, while keeping their main functionality.

\parrafo{Context} We assume that a company developing software for IoT devices has developed a software using JS on a Raspberry Pi (\textit{RPI}) or some superior device, \eg{} a regular PC, and must port it to more constrained devices, like SoCs. Because most SoCs devices are programmed using C, they would either translate their JS application to C or compile an embedded JS interpreter and transfer it with the JS application to the SoCs device. The first scenario implies the cost of maintaining two applications in parallel (one in JS and one in C), while the second scenario only requires to remove unnecessary features of the interpreter to run the application on the desired constrained devices. Hence, in this work, we focus on this second scenario.

\parrafo{Method} We address the problem of software miniaturization by formulating it as a multi-objective optimization problem, which considers interpreter size, memory usage, execution time, and company's device ranking. \momit~takes as input (1) the set of available code features $F$; (2) a list $L$ of compulsory features required by the program $ComF \subset F$; and, (3) a list of candidate IoT devices with their technical specifications (storage, memory). \momit~finds a set of combinations of optional features plus the identified compulsory features that best satisfy the devices' constraints for each objective.

\parrafo{Results} The results of this work are: 

\begin{enumerate}
\item We perform an empirical study to measure the impact of configuration options of a well-known JS engine (\duktape) on performance metrics. We identify, from a set of 286 features, 86 features that companies can activate/deactivate to improve the performance of their JS applications and potentially miniaturize them to run on more constrained devices.

\item We present a multi-objective miniaturization approach (\momit) that does not require to modify the JS source code, but the configuration options of the JS interpreter.

\item We implement \momit~on three different search algorithms, including NSGA-II, \hrs, and SWAY. We compare these algorithms in terms of execution time, and quality of solutions, to allow practitioners to make informed decisions.

\item We reveal 10 hidden dependencies within 20 of the 86 features studied that hinder the compilation and execution of JS interpreters when their default values are altered individually. 

\item We detect and report  a bug on the configuration of \duktape~options that prevents developer to customize their JS interpreter.  The bug was fixed and the corresponding patch committed to the repository's master branch by one of the authors of \duktape~(\textit{issue number 1990}). 

\item We conduct a comprehensive empirically study on 23 JS tests belonging to a benchmark for internet browsers and show that \momit~can port the tests to more constrained devices, the size of a quarter coin, after the miniaturization process.

\item We release \momit~and the three implemented algorithms as open source along with generated results, and raw data for the IoT community.

\end{enumerate}

Although we focus on the example of miniaturizing JS interpreters to constrained devices, our approach is general enough to support other code interpreters like Python, Lua, etc.

\subsection{Motivating Example}


A company specializes in the development of software applications for IoT.  Without loss of generality, we use JS as a the programming language used by that company. It wants to deploy its applications on a large set of IoT devices to attract a large customer population. To maximize its profit, it has decided to rank each device according to its potential market and--or existing customers' preferences for IoT devices like \textit{RPI},  \photon, \textit{ESP32}, etc. 

To run its applications on the more constrained devices, the company must \emph{miniaturize} the JS interpreter based on the features used by the JS code.  \emph{Miniaturization} is the process of removing unnecessary features in an application and--or its interpreter (or virtual machine) without modifying the original code of the application/interpreter~\cite{ali2011moms}. 

The company considers three internal attributes for each device: storage capacity, memory capacity and CPU power. Reducing storage size is important when porting applications to IoT devices with limited storage space. While Single-Board-Computers (SBCs), like \emph{RPI}, allow the use of microSD cards, which eventually can be expanded by a larger capacity card, SoCs are more limited in storage capacity (1 MB or less) and cannot be further expanded. Memory capacity cannot be expanded in any IoT devices and it is specially small on SoCs devices.

To miniaturize its application, the company could apply previous approaches like the one proposed by Ali~\ea~\cite{ali2011moms}, but they would not reduce the size of the interpreter that takes 314 KB of storage capacity, and 84 KB of memory using default JS interpreter configuration options~\footnote{Size of executing binary \duktape~JS interpreter on our lab. machine.}. Thus, the company decides to rather miniaturize the JS interpreter that will run their JS application. \momit~can serve to this purpose and only requires the following input: (1) the JS application to port; (2) application's pre-requirements (PRs); (3) the  list of preferred IoT devices to run the application with their corresponding technical specifications;  (4) a customizable JS engine that allows to (de)activate JS features on demand.


\textit{The Remainder of this Paper is Organized as Follows}. In \Cref{sec:relatedwork} we relate our work to the state of the art, while \Cref{sec:background} provides foundations on multi-objective optimization, evolutionary algorithms, the breadth and depth of IoT systems, JS, and software miniaturization, necessary to understand this work.
In \Cref{sec:approach}, we present our automated multi-objective approach for miniaturizing JS engines.  \Cref{sec:prelim} presents a preliminary study regarding the impact of JS engine features on performance metrics. In
 \Cref{sec:implementation} describes the implementation of our approach on different evolutionary algorithms. \Cref{sec:evaluation} describes the experimental setting for evaluating the proposed approach. \Cref{sec:results} presents the results obtained from our experiments, while \Cref{sec:momit-discussion} discuss the results obtained.
In \Cref{sec:threats-validity} we discuss the threats to the validity of our study.
Finally, we present our conclusions and highlight directions for future work in \Cref{sec:conclusion}.

\section{Related Work}
\label{sec:relatedwork}

In this section, we present relevant work. We divide the work on four categories.

\subsection{Programming Language Migration}
Programming language migration consists of developers manually porting the source code written in a programming language to other languages.  This is a tedious and error-prone task as it requires users to manually define migration rules between the source and the target code program constructs, including mappings between the equivalent APIs interfaces of third-party libraries used in the source code.  (Semi)automatic  tools have been proposed to mitigate the effort of migrating code.  For example,
Mossienko~\ea{}~\cite{Mossienko2003} proposed an approach to migrate COBOL code to C, or Sharpen~\footnote{https://github.com/mono/sharpen}, which allows developers to migrate Java code to C\#.  Other tools have been developed to support API migration of websites to modern APIs~\cite{Hassan2005,tonelli2010}.
The drawback of these tools is that they require developers to define migration rules to customize and perfect their conversion.  To the best of our knowledge, there is no tool support available for migrating JS code to C, which is the main programming language for IoT devices.

\subsection{Software Product Lines}
A software product line (SPL) is a collection of related software products, which shared core elements between them~\cite{harman2014search}.  From one single software product line, several products can be generated.  As example, consider the work of Apel~\ea~\cite{siegmund2012predicting} to model the compilation configuration options of database systems as a product line.  By tuning those configuration options according to a defined criteria,  different database products can be generated.  The great challenge of finding valid products using SPLs is that it becomes very difficult when the number of features increases.  In other words, random values of \textit{use} or \textit{not use} assigned to the features explored have low probability of satisfying the constraints of a valid product.  In this work, we corroborate this while implementing a \textit{pure random search} algorithm to test \momit, and found that not a single solution generated out of 250  was valid, due to the dependency between features discussed later in \Cref{sec:results}.
Recent works like Sayyad~\ea~\cite{sayyad2013scalable} and Chen~\ea~\cite{8249828} combine metaheuristics with a preprocessor (a SAT solver) to reduce the search-space to a subset of feasible solutions from the total space, instead of randomly generate solutions and evaluate them, which is typically done by evolutionary algorithms.

\subsection{Compilers Optimization}
Compiler flag selection can be an effective way to increase the quality of executable code according to different code quality criteria. 
Modern compilers can work on many platforms and implement a lot of optimizations, which are not always tuned well for every target platform. 

Hossein et al. \cite{DBLP:conf/date/AshouriPS16} evaluated different autotuning approaches including the use of Design Space Exploration (DSE) techniques and machine learning to further tackle the problems of selecting (choosing which optimizations to apply) and the phase-ordering (choosing the order of applying optimizations) of compiler optimizations. They demonstrated that these techniques have positive effects on the performance of applications and can bring up to 60\% improvement with respect to standard optimization levels on the selection problem and up to 4\% with respect to LLVM's standard optimization on the phase-ordering problem. Souza and Silva \cite{DBLP:journals/cai/XavierS18} presented a design-space exploration scheme, which aims to search for a compiler optimization sequence. The proposed hybrid approach relies on sequences previously generated for a set of training programs, with the purpose of finding optimizations and their order of application. In the first step, a clustering algorithm chooses optimizations, and in the second step, a metaheuristic algorithm discovers the sequence, in which the compiler will apply each optimization. They evaluated the approach using the LLVM compiler. The results showed that optimized sequences generated codes that outperformed the standard optimization level O3 by an average improvement of 8.01\% and 6.07\%, on Polybench and cBench benchmark suites, respectively. Plotnikov et al. \cite{PLOTNIKOV20131312} presented a tool for automatic compiler tuning, which helps to identify underperforming compiler optimizations. Using GCC for ARM, they showed how this tool can be used to improve performance of several popular applications, and how the results can be further analyzed to find places for improvement in the GCC compiler itself.
Luque et al. \cite{DBLP:conf/idc/LuqueA18} used parallel meta-heuristic techniques to automatically decide which optimization flags provided by the GCC compiler should be activated during the compilation on a set of programs. 
The proposed approach was able to adapt the flag tuning to the characteristics of the software, improving the final run times with respect to other spread practices.
P{\'{e}}rez et al. \cite{DBLP:conf/ae/CaceresPFS17} showed that the performance of compiled code has significant stochasticity, just as standard optimization algorithms. As a practical case study, they considered the configuration of the GCC compiler for minimizing the run-time of machine code for various heuristic search methods. Their experimental
results showed that, depending on the specific code to be optimized, the improvements of up to 40\% of execution time when compared to the -O2 and -O3 optimization flags is possible. Georgiou et al. \cite{DBLP:journals/corr/abs-1802-09845} also observed that by performing fewer of the optimizations available in a standard compiler optimization level such as -O2, while preserving their original ordering, significant savings can be achieved not only in execution time but also in energy consumption. 


\subsection{Software Miniaturization}

Software Miniaturization was first introduced by Di Penta~\ea~\cite{di2005language}, who proposed a software renovation framework to reduce the footprint of software during its porting to hand-held devices.  Their approach helps removing dead code, refactoring of code clones and the elimination of circular dependencies using clustering. They evaluated their framework on a geographic information system and claimed to reduced the average number of objects about 50\%.  They did not formulate their approach as multi-objective, and CPU time was not consider as an objective.

Ali~\ea~proposed \textit{MoMs}, a multi-objective approach for the miniaturization of applications based on user's prerequisites, storage occupation and CPU consumption~\cite{ali2011moms}.  They apply their approach to two applications (an email client and a instant messenger) and show that they could reduce the effort by 77\% on average, over a manual approach. Different from \textit{MoMs}, we take and indirect approach by miniaturizing the interpreter that executes an application, and not the application itself.  The advantage is that our approach does not modify the application in question.  Eventually both approaches could be applied on the same application to reduce the application footprint even more.

To the best of our knowledge, we are the first to address the problem of miniaturization of code interpreters for IoT using a multi-objective approach.


\section{Background}
\label{sec:background}

We now provide the necessary background on multi-objective optimization and IoT devices architectures to understand our proposed approach.

\subsection{Software Miniaturization}

\emph{Software bloat} is the proliferation of unused software components (features) that increases the size of binaries and libraries and affects software maintainability. In this work, we leverage the existence of \emph{software bloat} in JS interpreters to reduce their memory and code size footprints and to port them to constrained devices using the specific requirements of a JS application.

\subsection{Multiobjective Optimization}

The general formulation of a \emph{multi-objective optimization problem} is given by:

\begin{equation}
\begin{array}{rcl}
&\mbox{minimize }\ \ &\lbrace f_1(\mathbf x), f_2(\mathbf x), \dots , f_k(\mathbf x) \rbrace \\
&\mbox{subject to } \, &\mathbf x \in S,
\end{array}\label{prob}%
\end{equation}

\noindent where $k$ ($k \ge 2$) \emph{objective functions} $f_i:\mathbb R^n \rightarrow \mathbb R$ ($i = 1, \dots, k$) must be minimized at the same time and $S \subset \mathbb R^n$ is the \emph{feasible set}. A decision vector $\mathbf x=(x_1, \dots ,$ $x_n)^T$ is a \emph{feasible solution} if it belongs to the feasible set $S$. Its image $\mathbf z = \mathbf f(\mathbf x) = (f_1(\mathbf x), \dots , f_k(\mathbf x))^T$ is an \emph{objective vector} and the set of all objective vectors, denoted as $Z = \mathbf f(S) \subset \mathbb R^k$, is the \emph{feasible objective set}.

Usually, the conflict  between the objectives makes it impossible to find a feasible solution that simultaneously minimizes all of them. There is usually a set of Pareto optimal solutions on which none of the objectives can be improved without deteriorating, at least, one of the others. Given $\mathbf{z},  \mathbf{z'} \in \mathbb{R}^k$, we say that $ \mathbf{z}$ \emph{dominates} $ \mathbf{z'}$ if $z_i \leq z'_i$  for all $i = 1,\dots,k$  and $z_j < z'_j$ for, at least, one index $j$. When $ \mathbf{z}$ and $ \mathbf{z}'$ do not dominate each other, we say that they are \emph{non-dominated}. For problem \eqref{prob}, a \emph{Pareto optimal solution} is a feasible solution ${\bf{x}} \in S$ for which there does not exist another ${\bf{x}}' \in S$ such that $\mathbf{f} (\mathbf{x}')$ dominates $\mathbf{f} (\mathbf{x})$. The set of all Pareto optimal solutions in the decision space, denoted by $E$, is named the \emph{Pareto optimal set} and its image in the objective space, denoted by $\mathbf f(E)$, is called the \emph{Pareto optimal front} ($PF$).

\subsection{Multi-objective Evolutionary Algorithms}

\emph{Evolutionary Multi-objective Optimization} (EMO) algorithms have demonstrated their ability for solving multi-objective optimization problems \cite{Coello_EMObook,Deb_EMObook,Ishibuchi2008,Li2015,Caietal2017,Qietal2014}. They find a subset of non-dominated solutions approximating $PF$ (the set of all Pareto optimal solutions in the objective space). The approximation set is composed of solutions as evenly distributed as possible in the $PF$ (diversity) and as close as possible to the true $PF$ (convergence). A famous EMO algorithm is NSGA-II \cite{DebNSGAII}, which has successfully solved many real-life multi-objective optimization problems \cite{Deb_EMObook,Zhou2011}. It uses an elite-preserving strategy and a diversity preserving mechanism and it stands out by its fast non-dominated sorting procedure to rank the solutions into several non-dominated fronts for the selection of the best individuals.
 

\subsection{Breadth and Depth of IoT systems}

IoT applications involves machine-to-machine and human-to-human communications to deliver a variety of services to participants. IoT development road-map~\cite{iot_key_concepts} involves a range of key topic areas including: hardware, networking, application design, application development, security, business intelligence, data analytics, machine learning, and artificial intelligence. In this work, we focus specifically on hardware and application development.

\parrafo{Hardware} In the context of the IoT, a \emph{device} is an overloaded term that describes hardware designed/adapted to perform a particular task. In this work, we consider off-the-shell boards, which we divide in two categories: SBCs and SoCs devices. Generally, IoT devices are characterized by their data acquisition and control capabilities, data processing and storage capabilities, connectivity, and energy consumption. In \Cref{table:usr_paper}, we present some selected devices used in this work for comparison purposes. This is not a exhaustive list but a selection of the most relevant IoT devices available at the moment of writing this work according to IBM~\cite{iot_best_hardware}. Column ``Cloud enabled" indicates if the device includes pre-integrated cloud platform to manage a set of IoT devices. This feature is only available for \emph{Particle} devices. We did not include \emph{Arduino} devices because in majority they target hobbyists rather than industrial projects. For example, \emph{Arduino UNO} offers the lowest resources in terms of hardware (only 2 KB of memory and 32 KB of storage), does not include Wi-Fi, and has a price greater than \photon~and \emph{ESP32} ones.

\begin{table*}[ht]
\centering
\renewcommand{\arraystretch}{1.0}
\renewcommand{\tabcolsep}{.5mm}
\caption{List of IoT devices used in this work for comparison purposes}
\label{table:usr_paper}
\scriptsize
\begin{tabular}{|l|l|r|r|l|l|r|l|r|}
\hline
Name           & Processor                                  & Memory (KB) & Storage (KB) & Wi-Fi & Dimensions (mm)      & Weight (g) & Cloud enabled & Price (US) \\ \hline
\photon        & STM32 ARM Cortex M3                        & 128                   & 1,000                  & yes              & 36.58 x 20.32 x 4.32 & 5          & yes            & 19.00         \\ \hline
\textit{ESP32}                  & XTENSA DUAL-CORE-32-BIT                    & 512                   & 4,000                  & yes              & 55.3 x 28.0 x 12.3   & 9.6        & no             & 19.95      \\ \hline
\textit{JN5168} & \textit{JN5168} 32 bit RISC microprocessor          & 32                    & 256                    & no               & 24.5 x 30.5 x 9.77   & 4          & no             & 26.95      \\ \hline
\textit{RPI} 3 Model B+ & ARM Cortex-A53 CPU & 1,000,000             & 16,000,000             & yes              & 85 x 56 x 1.6        & 42         & no             & 54.40       \\ \hline
\textit{BeagleBone Black}       & AM3358 ARM Cortex-A8                       & 256,000               & 4,000,000              & yes              & 86.40 x 53.3        & 39.68      & no             & 89.00         \\ \hline
\end{tabular}
\end{table*}


\parrafo{Software} The standard programming language for developing applications on IoT devices is C/C++, without or with IDE support, among which \emph{Arduino} IDE is one of the most popular; others development environments support JS, like \emph{Tessel} and \emph{Particle.io} while \emph{MicroPython} and \emph{WeIO} support Python. Developers opt for cross-platform IDEs like \emph{Arduino} IDE to mitigate the burden of developing/using device-specific libraries for each different device that they target. However, there are physical differences that they still must be considered, for example the voltage of digital I/O pins vary from device to device (5 Volts on \textit{RPI} to 3.3V on SoCs). 

\subsection{JavaScript}

JavaScript (JS) is a high-level interpreted programming language, (\ie~requires an interpreter to execute code line by line, in contrast with compiled programming languages whose code is translated to an intermediate machine language before being executed). Because SoCs are constrained by memory and storage capacity, deploying a full JS interpreter is not feasible. Although there are devices that can run JS code natively (\ie~ Espruino), this is not an applicable solution for all existing devices.  Alternatively, JS engines for embedded devices can serve to compile lightweight JS interpreters to deploy applications in highly-constrained environments (\eg~\duktape~\cite{duktape}, tinyJS\footnote{\url{http://tinyjs.net}} and JerryScript\footnote{\url{http://jerryscript.net}}). Still,  developers must adjust their code to use the APIs of the selected JS interpreter, requiring additional effort. In this work, we use \duktape~to produce a miniaturized JS interpreter. The rationale of this choice are its portability and compact footprint and customization APIs, which are ideal for embedded devices constrained in memory and CPU capacity. Because \duktape~allows to build an embedded JS interpreter in C/C++, it reduces conflicts with the internals of the hardware architecture. \duktape~does not provide printing or Input/Ouput (IO) capabilities, but instead allows JS code to communicate with native IO functions in C. \duktape~provides bidirectional communication means between C and JS functions and it is possible to optimize even more the performance of code functions that, otherwise, would be slower if they were implemented in pure JS.

\subsubsection{ECMA Script}

ECMAScript (ES) is a trademarked scripting-language specification standardized by Ecma International in ECMA-262 and ISO/IEC 16262\cite{ecmascript}. It was created to standardize JS to foster multiple independent implementations and prevent fragmentation. ES extends from ES version 5 to version 9. In ES~5, we find array, date, and math built-ins, while later standards enable support for more complex features, like reflection (ES~6), or encoding built-ins for UTF-8 (ES~9). 






\subsection{\duktape{}}
\label{subsec:JSinterpreter}

\duktape{} is an engine to embed JS interpreters on C/C++ code, though it can also be used as an standalone executable interpreter.  In this work, we use the terms configuration options and features indifferently. The main difference is that configuration options are provided by \duktape~to control JS interpreter's functionality, while features are the functionality of interest.  For example, array's built-in. 

According to the official \duktape{}'s website\cite{duktape}, it is possible to execute \duktape{} in constrained environments with a minimum of 160 KB of storage and 64 KB of memory.

We exemplify the use of \duktape{} as an embedded interpreter in \Cref{lst:harness}, which we called \emph{Harness}. \emph{Harness} takes as parameters the name of a JS file and the name of a JS function to be executed. From Lines 6 to 18, there is a static function to push the JS file into \duktape's \emph{context}. \emph{Context} is an ES execution thread, which resides in a \duktape~heap. The heap is a memory region used to allocate storage for strings, ES objects, etc.\ for garbage collection. The \emph{call} stack registers the active function call chain of a context and the \emph{value} stack stores values belonging to the current activation in a \emph{context's call} stack. Values kept in a \emph{value} stack are tagged types (\eg~boolean, string, object, etc.). Stack entries are indexed either from the bottom (negative values) or from the top (positive values) of the most recent function call.

From Lines 19 to 25, there is a static function to compute the amount of memory used by the execution of the JS code. In this work, we use Linux \emph{mallinfo} command to obtain information about the memory usage, specifically memory allocations performed by \emph{malloc} and related functions. We are interested in the total allocated bytes (\emph{uordblks})

Lines 26 to 62, correspond to the main function of \emph{Harness}; in Lines 28 to 34, \duktape{} heap is initialized. Lines 36 to 40, we push the JS file as a string into the \emph{heap} and evaluate it. At this line, no JS code is executed yet. In Lines 44 to 51, the JS function specified as second parameter is called or the complete JS file is executed if the character ``." is found. Next, memory used after executing the JS file is measured (Line 52).

Finally, the \emph{heap} is destroyed and memory allocated is released (Line 55).  

\begin{lstlisting}[,language=C, caption=Example of \duktape{} as embedded JS interpreter in C code,label={lst:harness},escapeinside={(*@}{@*)},basicstyle=\scriptsize,numbers=left,xleftmargin=15pt,breaklines=true]
/ *  This script reads JS scripts and measures memory footprint. Parameters: JS file, main JS function to execute  */
// we ommit includes of standard C libraries due to space constraints
#include "duktape.h"
/* The maximum size of the JS file to read.  This value is adjusted based on the size of the file to execute. */
static void push_file_as_string(duk_context *ctx, const char *filename) {
    FILE *f;
    size_t len;
    char buf[17384];
    f = fopen(filename, "rb");
    if (f) {
        len = fread((void *) buf, 1, sizeof(buf), f);
        fclose(f);
        duk_push_lstring(ctx, (const char *) buf, (duk_size_t) len);
    } else {
        duk_push_undefined(ctx);
    }
}
static int
       get_total_size_of_memory_occupied(void)
       {
           struct mallinfo mi;
           mi = mallinfo();
           return mi.uordblks;
       }
int main(int argc, const char *argv[]) {
	if ( argc == 3){
	    duk_context *ctx = NULL;
	    (void) argc; (void) argv;
	    ctx = duk_create_heap_default();
	    if (!ctx) {
	        printf("Failed to create a Duktape heap.\n");
	        exit(1);
	    }
        /* We push to Duktape heap the name of the JS file*/
	    push_file_as_string(ctx, argv[1]);
	    if (duk_peval(ctx) != 0) {
	        printf("Error: %s\n", duk_safe_to_string(ctx, -1));
	        goto finished;
	    }
	    duk_pop(ctx);  /* ignore result */
	    /* when the js function argument is different from 0, 
		*  we should call the function specified by the argument.*/	
	    if (strcmp(argv[2],".")!=0){
			duk_push_global_object(ctx);
	    	duk_get_prop_string(ctx, -1, argv[2]);
	    }
	    if (duk_pcall(ctx, 0 /*nargs*/) != 0) {
	        	printf("Error: %s\n", duk_safe_to_string(ctx, -1));
	            } 
	    duk_pop(ctx);  /* pop result/error */
	    int JS_MemoryUsed = get_total_size_of_memory_occupied();
	    printf("%d",JS_MemoryUsed );
		finished:
	    duk_destroy_heap(ctx);    
	}
	else {
		printf("You need to provide: (1) the name of the JS to run;\
			(2) the JS function to execute"); 
	}
	exit(0);
}
\end{lstlisting}



\section{Approach}
\label{sec:approach}

We now introduce \momit~(Multi-objective Software Miniaturization for the Internet of Things). We describe in detail each step of \momit~based on the motivating example presented in~\Cref{sec:intro}.  We describe the \momit~process as it would be applied by the company without making any assumption on the available tool support, while in \Cref{sec:implementation} we provide details of how to implement \momit~.

\subsection{Pre-requirement Elicitation}

The pre-requirement elicitation step, consists of determining the set of JS interpreter features required to execute their code, including the compliance to a particular ES  standard. The features can be determined by asking the authors of the applications and--or by using code static-analysis tools, like JSAnalyse~\footnote{https://archive.codeplex.com/?p=jsanalyse}.

\subsection{Selection of IoT Device Candidates}

In this step, the company must provide the list of IoT devices on which it would like to deploy its applications. This list includes the technical specifications (\eg~memory and storage capacity) and rank number for each device based on the company's preferences.  For example, if the company prefers to deploy its applications on a specific device (\eg~\photon), then solutions that fit \photon~devices will be prioritized by \momit.

\subsection{Feature Identification}

Next, developers map each PR to one or more features of the JS interpreter. \momit~receives as an input a list of JS features $F$ with their corresponding dependencies $DepF$. A feature $f_i \in F$ has a dependency with a feature $f_j \in F$ with $i \neq j$, if $f_i$ requires $f_j$ to have a value $v \in \{false,true\}$  to produce a valid JS interpreter. In this work, a valid JS interpreter is a customized JS interpreter derived from a set of selected features based on a PR analysis, and that was successfully build.

Additionally, the aforementioned company must supply a list of compulsory features $ComF$, which is a subset of $F$, in case there are features required to execute the application.  In the case that a company requires to comply with specific version of ES, they can include those features in $ComF$ to prevent \momit~for deactivating those features.  It is also important to mention that there also exist features related to the basic behavior of the JS interpreter, and which we cannot disable, for example: \duktape's~heap, context,  arithmetic operator, primitive types, etc., and consequently are not considered in $ComF$.


\subsection {Selection of Feature Combinations}

Then, \momit~can determine a set of features satisfying application requirements  within the constraints imposed by the IoT devices' constraints. It starts with the compulsory features $ComF$ and completes them with a set of optional features $OF \equiv \left\{g_1, \ldots, g_{N}\right\}$ where $N$ is the number of optional features and $g_i \in F$ with $i=\lbrace 1,\dots,N \rbrace$., by performing multi-objective optimization. A miniaturized interpreter can implement $F'\subset OF$ optional features and there exist $2^{OF}$ possible sets $F'$. 

\momit~considers that an interpreter is comprised of $M$ implementation units $IU\equiv\left\{iu_1,iu_2\ldots,iu_M\right\}$. Function $Impl$ is a function that takes as input a set of features and returns the corresponding implementation units. We define a miniaturized interpreter as $IU'=Impl(F'\cup ComF)$ where $F'$ is the  set of selected optional features and $ComF$ of the compulsory features. 

Including/excluding a feature requires dealing with a set of property values $P\subset\mathbb{R}^K$  concerning the device usage (code size, memory usage,  execution time), and with a set of internal constraints $IC\equiv\left\{ic_1,ic_2,\ldots,ic_K\right\}$, each of them imposing a set of $ic_j$ acceptable values on the corresponding property values, with $P\in IC \equiv\left\{p_j \in ic_j \forall j=1, \ldots, K\right\}$. Function $Prop_j(IU')$ returns the property value of an interpreter with respect to constraint $j$ with $j \in IC$. For the sake of simplicity, we focus on code size, memory usage, and execution time, although other constraints can be considered (\eg~network connectivity, energy consumption, device form factor, etc.).  We represent the set of potential IoT devices to port an interpreter as $L\equiv \left\{l_1, l_2 \ldots,l_l\right\}$.

To measure to which extent a miniaturized program $IU'$ matches the internal constraints of a device $l$, we define the device satisfaction rate of $l$ in Equation~\ref{eq:dsrofl}:

\begin{equation}
\label{eq:dsrofl}
DSR_l(IU') = \frac{ \sum_{j=1}^{K} \frac{Prop_j(IU')-ic_j(l)}{ic_j(l)} }  {K} 
\end{equation}

The company (from the motivation example) ranks each device $i$ according to its value based on its potential market and--or existing customers' preferences: $val_i$, where $1 \leq val_i \leq V_{max}$ and $Val \equiv{val_1,val_2,\ldots,val_L}$. We use $Val$  to define an overall satisfaction measure,  or customer's satisfaction rate (USR), as presented in Equation~\ref{eq:USR}. USR expresses the extent to which a solution (\ie~miniaturized interpreter for IoT) matches the customer's device preferences.

\begin{equation}
\label{eq:USR}
USR(IU') = \frac{ \sum_{i=1}^{L} DSR_i(IU')\times \frac{val_i}{V{max} }} {L}
\end{equation}

The formulation to solve the problem of miniaturizing an interpreter for the IoT is presented in Equation~\ref{eq:fitnessfunction}. By solving Equation~\ref{eq:fitnessfunction}, \momit~obtain optimal combinations of features, which are miniaturized JS interpreters, implementing subsets of the features of a JS interpreter using default features values, so that they: (1) maximize customer's device preferences $USR$, \ie~minimize costumer's dissatisfaction device rate $-USR$, and (2) satisfy the constraints IC by minimizing a set of property values, $Prop(IU') \in IC$.

\begin{equation}
\label{eq:fitnessfunction}
\min_{F'\in2^{OF}}(-USR(IU'), Prop(IU'))
\end{equation}

We formulate the problem of miniaturization as a multi-objective problem and, thus, \momit~is likely to find more than one solution. A solution $F'$ is a set of features from $F$ that serves to build a JS interpreter able to execute a JS application provided by a company. Then, developers can use other criteria to select one single solution: for example, the solution that can be deployed in most devices; the solution that executes faster, etc.

\section{Preliminary Study}
\label{sec:prelim}


In this section, we describe the methodology that we followed to identify the JS interpreter features that \momit~will consider to include/exclude according to the pre-requirement elicitation and feature property analysis of the JS interpreter to be miniaturized.


We formulate the following research question:

\begin{myenumi}
\item[\bf{(PQ1)}] \PQone{} The rationale of this question is to determine the impact of each of the selected JS interpreter features. We test the following null hypothesis: \emph{$H_{0_1}$: there is no difference between the performance of the JS interpreter before and after modifying the default value of a selected feature}. 

\item[\bf{(PQ2)}] \PQtwo{} The rationale behind this question is to identify, from the measurements obtained in \textbf{PQ1}, which features report the higher values on the performance metrics studied. We test the following null hypothesis: \emph{$H_{0_2}$: there is no JS interpreter feature that is a major concern on performance metrics}. 
\end{myenumi}

We distinguish between two broad categories of features: (1) ES compliant-features and (2) JS interpreter-specific features. The first category includes all ES features from version 5 to 9, as currently supported by any JS interpreter.



We consider all ES compliant-features and explored the existing \duktape{} documentation in search of keywords related to performance. \duktape{} contains 286 features from which only 40 corresponds to ES features. Moreover, the values that we can apply to these features are not limited to a two-value nominal scale (activate/deactivate) but include ordinal, interval, and ratio scales to tune certain features, like the size of the debug-code static buffer. Studying each feature one by one is a time consuming task. Theoretically, it is even impossible given ratio scales. To shorten the search, we rely on \duktape{} configuration profile files located on the \texttt{config/examples} folder of the interpreter. The configuration profiles are templates for different application environments. For example for low-memory-, performance-sensitive-, timing-sensitive-, security-sensitive-environments, etc.  By analyzing the values assigned to each feature in these profiles, we can determine adequate values for features that requires ratio scales to reduce memory usage, or to improve execution performance.

Based on the information gathered from the documentation, and the configuration profile files, we identified \textbf{86 features} to measure, dividing into 40 ES-compliant features and 46 features related to performance metrics. We choose the following criteria to filter features: deprecated features (based on the existing documentation); features enabling extra debug features, as these features will only add overhead; features under development,  and experimental features, typically disabled by default.

To study the impact of the JS interpreter features, we developed a framework  to execute a JS application and measure its  performance (more precisely,  interpreter plus application) after (de)activating each of the 86 identified features one by one, except for the features that are depedent on other feature values to work. We wrote this framework in Python, leveraging \duktape{} Python configuration script, which produces C source code and headers to embed the JS interpreter in a C program, based on a configuration file. 

Our framework is comprised of three Python scripts to support our study. The first script $p1$, benchmarks a JS application using the default \duktape{} features (generated also by \duktape{} configuration script). The second script $p2$ benchmarks a JS application, changing the default value of each of the 86 features individually for a value suggested either in some configuration profiles or in \duktape{} documentation to achieve certain goal (reduce code size, memory usage, etc.). The file with the complete list of features and their values can be found at the online replication package~\cite{MoMItRep}. The third script $p3$ benchmarks a JS application using the features and values read from a given text file. The C file with the embedded JS interpreter to perform the experiment is already introduced in \Cref{sec:background}, \Cref{lst:harness}.

Without loss of generality, we perform the measurements in a \emph{RPI 3} Model \emph{B+} with a 1.4GHz 64-bit quad-core \emph{ARM Cortex-A53} CPU, using \duktape{} 2.3.0, and \emph{gcc 6.3.0} compiler.


In \Cref{lst:primeSimple}, we provide the JS code~\cite{MoMItRep} that we wrote to benchmark each JS feature considered. \Cref{lst:primeSimple} implements an algorithm to count the number of prime numbers below $100,000$.  This is a minimal version that does not make use of strings, arrays, objects, or any libraries, but primitive types and standard arithmetic operators, so it is compatible with all features. In general, the performance metrics measured from executing \textit{primeSimple} have to be considered as an upperbound of how much improvement can be attained when changing a default feature value. We did not include any print statement as \duktape{} does not provide print built-ins.  Hence, any output redirected to screen must be implemented in C code and linked with \duktape{}.

\begin{lstlisting}[,language=Java, caption=JS source code used to benchmarking JS interpreter selected features,label={lst:primeSimple},escapeinside={(*@}{@*)},basicstyle=\scriptsize,numbers=left,xleftmargin=15pt,breaklines=true]
function isPrime(num) {
  for(var i = 2; i < num; i++)
    if(num % i === 0) 
      return false;
  return num !== 1;
}

function forTest() {
  count =0;
  for (var i = 1; i < 100000; i++) {
    if (isPrime(i) && (i % 10000) == 9999) {
      count++; 
	}
   } 
}

\end{lstlisting}

Algorithm~\ref{algo:p2} ($p2$) presents the steps used for benchmarking individually each of the selected JS features for this study. For sake of simplicity, we did not present $p1$ and $p3$ algorithms, because they have slight variations with respect to $p2$. All the scripts used in this experiments can be downloaded from the online replication package of this work~\cite{MoMItRep} .

\begin{algorithm}[ht]
\DontPrintSemicolon
\SetKwInOut{Input}{Input}
\SetKwInOut{Output}{Output}
\Input{$S$, $harness$, \textit{file with features},  $codeSizeJ$, $memUsJ$, $execTimeJ$, $runs$ }
\Output{$report\_file$}
$report\_file=\emptyset$\;
 	\ForAll{row $\in$ file with features}
 	{
    	Save feature and its test value on ($cfg$)\;
        Generate $j'$ based on $cfg$ \;
        Compile $harness$ with $j'$ \;
        \If {compilation fails} 
        {Continue with next $row$ \;} 
        $codeSizeJ'$ = measure code size ($harness$)\;
        $\delta codeSize$ =
            $\frac{codeSizeJ'-codeSizeJ}{codeSizeJ}$ \;
        Open report file\;
 		\For{$1$ \KwTo $runs$}
 		{
            Execute $harness$ with $S$\;
  			$memUsJ'$ = measure memory usage ($harness$)\;
            $\delta memUs$ = $\frac{memUsJ'-memUsJ}{memUsJ}$ \;
            $execTimeJ'$ = measure execution time ($harness$)\;
            $\delta execTime$ = $\frac{execTimeJ'-execTimeJ}{execTimeJ}$ \;
            Write $\delta codeSize,  \delta memUs, \delta execTime$ to $report\_file$\; 
		}
        Close $report\_file$\;
         \KwRet $report\_file$\;
	}
    \KwRet $report\_file$\;
 \caption{Steps to benchmark the selected JS interpreter features.}
 \label{algo:p2}
\end{algorithm}

Algorithm~\ref{algo:p2} receives as an input a JS file ($S$); a C file with an embedded JS interpreter ($harness$), which is the code that we miniaturize and port to a constrained device; a file containing JS interpreter's features with the values we are interested to test (\textit{file with features}), and the performance metric values measured by executing $S$ using default \duktape~features ($j$); and the number of time to execute $harness$. Since execution time and memory usage may vary from run to run due to the non-deterministic way that CPUs and operating system task schedulers work, we set $runs=10$ to control for possible variations of measurements.

A $row$ in \emph{file with features} (Line 2) represents a pair \{feature name, test value\}. In Lines 3 to 4, we use \duktape{} configuration script (\cf~\Cref{sec:background}) to generate a new JS interpreter $j'$ by reading a configuration file (\emph{cfg}) generated from $row$. For example, to remove the default array built-in for the JS interpreter, we will write \texttt{DUK\_USE\_ARRAY\_BUILTIN:FALSE} in \emph{cfg}, with \texttt{DUK\_USE\_ARRAY\_BUILTIN} the name of the feature to (de)activate arrays.

In Line 5, we compile $harness$, which is the C file needed to read and execute $S$. If the compilation fails, then we skip to the next feature. In general, the compilation of $harness$  fails if a feature required to run $S$ is not present or in case of existing dependencies among features. For example, from \duktape{} low-memory profile, we learned that for storing strings in the ROM of the target device, we need to set 4 feature values to True.  For sake of simplicity, we treat features that have to be (de)activated together as one group of features.  Non-documented dependencies among features hinder the process of feature selection.  We called \emph{non-documented dependencies} those dependencies who are not documented either in the API of \duktape{}, or in the profile examples; for those features it is necessary to go to the source code of the interpreter to find them. For example, the dependency between feature \texttt{DUK\_USE\_JSON\_BUILTIN} (26) and \texttt{DUK\_USE\_JSON\_SUPPORT} (27).  Feature 26 requires feature 27 to be activated (true) in order to build a valid JS interpreter, but this is not mentioned in the documentation of \duktape{}. We present the complete list of  \emph{non-documented dependencies} found in this preliminary study, and in the evaluation of \momit~in \Cref{sec:evaluation}.

In Line 9, we measure the size of the compiled interpreter ($codeSizeJ'$) using the Linux command \texttt{stat}, which reports the number of bytes of the miniaturized JS interpreter, and in Line 10, we compute the difference of code size using default and the test value.  In Line 11, we open a report file to store the performance metrics of the multiple runs executed.  $codeSizeJ'$ does not change between runs.  Hence, we measure it only once.

Lines 12 to 19 , we execute $harness$ passing $S$ and its corresponding parameters.  We measure memory usage using the same procedure explained in \Cref{subsec:JSinterpreter}. To measure execution time we use Linux's \texttt{time} command (not the Bash time command, but the one located at \texttt{/usr/bin/} directory), and we report the total number of seconds that the process spent in user mode.

The output of Algorithm~\ref{algo:p2} is a CSV file with the percentage change ($\delta$) of each performance metric ($pm \in PM$), defined in \Cref{eq:delta_m}: 

\begin{equation}
	\label{eq:delta_m}
    \delta(pm) = \frac{\text{median}(pm(J'))-\text{median}(pm(J))}{\text{median}(pm(J))}
	\end{equation}

Where negative values indicates an improvement in $pm$ value, and positive values a detriment.

\subsection{Data Analysis}
In the following we describe the dependent and independent variables of this preliminary study, and the statistical procedures used to address each research question. 

\textbf{(PQ1):} \PQone\\
For \textbf{PQ1}, the \emph{dependent variables} are the measured performance metrics for each JS interpreter feature. The \emph{independent variable} is the use of default/test values on each JS-interpreter features, which can be dichotomy or continues values based on \duktape{} configuration files.  In total we found 75 dichotomous variables and 11 continuous variables.  We transform the continuous variables to dichotomous values, by considering the default value, and the value optimized in the configuration profiles exclusively.

\textbf{(PQ2):} \PQtwo\\
For \textbf{PQ2}, we analyzed the features corresponding to the outliers obtained in \textbf{PQ1}. The \emph{dependent} and \emph{independent} variables are  the same than in \textbf{PQ1}.

\subsection{Results and Discussion of the Preliminary Study}

\begin{figure}[ht]
\centerline{
\includegraphics[scale=0.3]{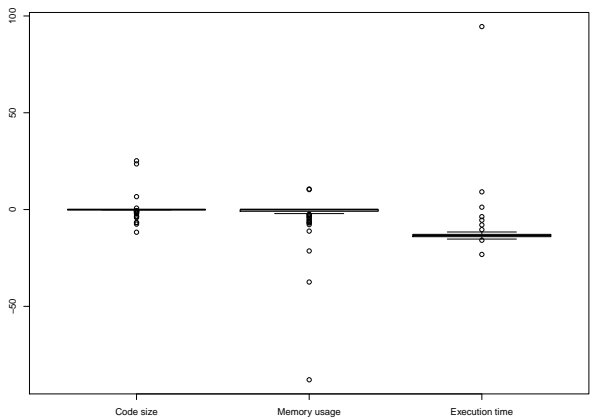} 
}
\caption{Percentage change in performance metrics for the 86 JS-interpreter selected features}
\label{fig:boxplot_prelim}
\end{figure}

In~\Cref{fig:boxplot_prelim} we present the distribution of percentage changes in performance metrics for the 86 JS-interpreter selected features. This value is calculated using \Cref{eq:delta_m}.  We observe that the boxes are flat in the three performance metrics studied, indicating that there is little variation of percentage change values among features.  We count the number of features measured different than zero, for each performance metric as: code size 52, memory usage 35, and execution time 82 respectively. That is at least 40\% of the selected features have an impact in any of the performance metrics studied. On the other hand, we did not find any feature where the percentage change is equal to zero for the three performance metrics studied at the same time.  Hence, we reject \emph{$H_{0_1}$}.

With respect to the magnitude of the impact, the medians of the percentage change for \emph{code size} and \emph{memory usage} metrics is small ($-0.01\%$, $0\%$), indicating that modifying the default values of a JS-interpreter features has little impact on code size and memory usage performance for most of the features studied.  On the other hand, the median percentage change value for \emph{execution time} goes down to $-13.41\%$ indicating that modifying the default values of the selected features positively improves the execution time in most of the cases.  We also observe that there are many outlier features, for each performance metric that are worthy to studyin more details in the next research question \textbf{PQ2}.

\hypobox{Overall, our results show that  JS-interpreter features impact performance metrics differently; that all features studied affect at least one performance metric, and execution time is the performance metric most improved} 

 From~\Cref{fig:boxplot_prelim}, we observe that there exist some outlier features that impact the performance of the metrics studied in a considerable manner.  The studied IoT devices, specially SoCs, are extremely constrained in memory and storage capacity, so giving the large amount of features and for improving readability, we focus the discussion on features that have an impact greater than $5\%$, and we called them features of major concern.
For code size we found the following percent values: -11.7, 25.2 and 23.6; for memory usage: -87.93, -37.39, -21.39, -11.13, -7.79, -6.79, -5.43, 10.32, 10.62.  For execution time, there are \textbf{78} features that impact more than 5\% the execution time of the prime simple JS, from which the lower values (higher improvement) are: -15.85, -23.17 and only two values that worsen the execution time: $9.15\%$ and $94.5\%$.  Since there are several features that impact performance metrics more than $5\%$ for each performance metric studied we reject \emph{$H_{0_2}$}.

In \Cref{table:prelim_major_concern} we present the features with the highest impact on performance metrics in our study.  Due to space limitations, we only present the two highest improvement features for execution time metric.  The complete list of features and performance metrics can be found in the article's replication package~\cite{MoMItRep}.  

\Cref{table:prelim_major_concern} is comprised of the following columns: \textit{id} is an arbitrary number that we assign to each feature for convenience; \textit{name} is the configuration option name, and \textit{description} is a summary of the feature functionality based on \duktape{}'s documentation'; \textit{value} is the tested value in our experiments (opposite from \duktape{}'s default value); and finally the percentage changes of three performance metrics measured.    

For code size, feature \textit{3} reduces code size $\approx$$12\%$ by increasing executing time around $9\%$; feature 11 reduced code size  $7\%$, and execution time $12\%$ while increasing memory usage $\approx$$11\%$.  These are the features that reduce the most code size.  On the other hand, using ROM's built-ins (features 7-10) increase code size $25\%$, and feature 86 $\approx$$24\%$. In general we observe that the reduction of code size it for each individual features is limited compared to the other metrics.

With respect to memory usage, the use of ROM's built-ins reduced it about $88\%$.  This is the highest reduction for all features and  performance metrics studied. The next two features that reduced the most memory usage are 5, 43 and 19 with an improvement of 37, 21 and 11 percent.  On the other hand, feature 11 increased memory usage by $11\%$ and it is the only feature that increases memory usage more than $5\%$ and it is related to the use of automatic reference counting for garbage collection.

With respect to execution time, the features with highest reduction are 84 and 15, with 23, and 16 percent. On the other hand, feature 6 reports the highest worsening of all features and metrics with 94\%, with a negligibly reduction for code size and memory usage.  Hence, practitioners should be aware of keep this function deactivated.

In general, we observed that the combination of different features affect differently the performance of a JS software. It is clear that the (de)activation of certain features will result in a high improvement of a metric while worsening one or more metrics.  In fact we noticed that only feature 5 presents not conflict with the metrics studied, as it  improves memory usage and execution time, while not affecting code size, but only affecting compliance with ES standard.

We selected the most basic example (no arrays, strings, encoders, etc.) to quantify as much features as possibly individually.  However, we also acknowledge that depending on the functionality of a JS software, the (de)activation of of these features could improve/worsen even more than the values presented here.  Additionally, we cannot estimate to what extent these percentage changes accumulate when different test values are applied at the same time, or if there is a threshold when values cannot be improved/worsen more.

Finally, we believe that this information is useful for practitioners, as help them to take informed decisions on the possible impact of changing the default values of the JS interpreter when porting their JS code to more constrained devices.

\hypobox{Overall, our results emphasize the existence of conflict between the three performance metrics studied.  There exists a need of providing practitioners with an automated approach to select the features that are more advantageous for the devices they are targeting, preferably a multi-objective approach where all performance metrics are non-dominated between each other.}

\begin{table*}[t]
\centering
\renewcommand{\arraystretch}{1.0}
\renewcommand{\tabcolsep}{.5mm}
\caption{JS-interpreter features of major concern for miniaturization to constrained devices}
\label{table:prelim_major_concern}
\scriptsize

\begin{tabular}{|p{0.05\textwidth}|p{0.24\textwidth}|p{0.5\textwidth}|p{0.05\textwidth}|p{0.05\textwidth}|p{0.05\textwidth}|p{0.05\textwidth}|}
\hline
id      & name                                & description                                                                                                                                                                                                                                                                                  & value & code size $\delta$ & mem. us. $\delta$ & exec. time $\delta$ \\
\hline
3       & DUK USE EXEC PREFER SIZE        & Prefer size over performance in bytecode executor                                                                                                                                                                                                                                              & TRUE  & -11.71      & 0          & 9.15                \\
7 to 10 & Divers (Store objects in ROM)       & built-ins for compiling objects and strings as constants in the ROM space to reduce RAM usage at the cost of a larger code footprint and slower performance                                                                                                                                                                                  & TRUE  & 25.21       & -87.93     & -13.41              \\
86      & DUK USE REGEXP CANON WORKAROUND & Use a 128kB lookup table for improving RegExp processing performance at the cost of code size                                                                                                                                                                                           & FALSE & 23.59       & 0          & -14.63              \\
5       & DUK USE LIGHTFUNC BUILTINS       & Force built-in functions to be lightweight functions. This reduces memory footprint by around 14 kB at the cost of some non-compliant behavior.                                                                                                                                              & TRUE  & 0           & -37.43     & -12.805             \\
43      & DUK USE BUFFEROBJECT SUPPORT     & Enable support for Khronos/ES~6 typed arrays and Node.js Buffer objects. This includes all ArrayBuffer, typed array, and Node.js Buffer methods.                                                                                                                                              & FALSE & -4.18       & -21.39     & -14.02              \\
19      & DUK USE DATE BUILTIN             & Provide a Date built-in                                                                                                                                                                                                                                                                      & FALSE & -0.11       & -11.13     & -13.41              \\
11      & DUK USE REFERENCE COUNTING       & Use Automatic Reference Counting for Memory management to remove objects that are no longer needed                                                                                                                                                                                           & FALSE & -6.79       & 10.62      & -12.2               \\
84      & DUK USE FASTINT                   & Enable support for 48-bit signed "fastint" integer values. Fastints are transparent to user code (both C and Ecmascript) but may be faster than IEEE doubles on some platforms. The downside of fastints is increased code footprint and a small performance penalty for some kinds of code. & TRUE  & 6.67        & 0          & -23.17              \\
15      & DUK USE ARRAY BUILTIN            & Provide an Array built-in.                                                                                                                                                                                                                                                                   & FALSE  & -1.7        & -4.85      & -15.85              \\
6       & DUK USE PREFER SIZE              & Catch-all flag which can be used to choose between variant algorithms where a speed-size tradeoff exists (e.g. lookup tables).                                                                                                                                                               & TRUE  & -0.76       & -0.03      & 94.51          
\tabularnewline
\hline
\end{tabular}%
\end{table*}

\section{Implementation}
\label{sec:implementation}

We describe the techniques and tools used for implementing \momit~to miniaturize JS interpreters for IoT constrained devices. Without loss of generality, we focus on \duktape{} as the JS engine to generate the miniaturized JS interpreters.

\subsection{Pre-requirement Elicitation}

As described in \Cref{sec:approach}, a miniaturized JS interpreter is comprised of compulsory and optional features. The list of compulsory features can be inferred through a manual inspection of the JS code and--or by asking the authors to provide pre-requirements, or by using a JS parser, and--or feature detection tools. The optional features form the search space of the problem and may differ between JS interpreters.

\subsection{Selection of IoT Device Candidates}

We used the devices already introduced in~\Cref{table:usr_paper} as the IoT device candidates used in this work. We rank them to set the preference that \momit~will use to filter solutions. In~\Cref{table:devices_preferences}, we arbitrarily set the preferences of devices based on  price, size, cloud connections, and company support of the candidate devices, from which \photon~is the more attractive, while BeagleBone Black is the less attractive due to its price, size, and lack of cloud support. We did not consider the processing power or memory capacity, as we assume that if a JS application can be miniaturized for the SBCs, it will be fit as well on the SoCs devices, so does not make senses to prioritize the miniaturization to fit them.

\begin{table}[ht]
\centering
\caption{List of IoT devices used in this work for comparison purposes}
\label{table:devices_preferences}
\begin{tabular}{ll}
\hline
Device name & val \\ \hline
\photon & 5 \\ 
\textit{ESP32} & 4 \\ 
\textit{JN5168} & 3 \\ 
\textit{RPI} 3 ModelB+ & 2 \\ 
\textit{BeagleBone Black} & 1 \\ \hline
\end{tabular}
\end{table}

\subsection{Feature Identification}

Another contribution of this work is the classification of relevant \duktape{} interpreter options in four categories based on their impact on performance and compliance with ES standards. This classification is valid for any JS interpreter.

\subsubsection{ECMAScript (ES) Compliance Options}

Features in this category provide built-in functionality to comply with ES standards. Including/removing features in this category affects the level of compliance of a JS interpreter.  Executing JS applications in an interpreter that does not comply with the specific ES for which the application was written for, may result in failure or unexpected behaviour. For example, ES6 requires to accept \textit{HTML} commenting style, and this feature is not available on ES5.

\subsubsection{Code Size Options}

Features in this category reduce code size of the JS interpreter at the cost of extra functionality of the JS interpreter. For example, disabling Unicode support for non-BMP characters reduces code size.

\subsubsection{Memory Performance Options} 

Features in this category relate to techniques to reduce the memory footprint of the JS interpreter at the cost of CPU speed and--or code size. For example, compiling objects and--or strings as constants and storing them in read-only memory (ROM) reduces startup RAM usage considerably at the cost of code size and slower performance.

\subsubsection{CPU Performance Options}

Features in this category relate to techniques aimed to reduce the CPU time of the JS interpreter at the cost of memory and--or code size. For example, using a look-up table to convert an object to string using \texttt{JSON.stringify} function to improve performance.

While in \duktape{} documentation, features are classified using a broader category schema that also includes debug options and \duktape{} specific options, we focus exclusively on those related to compliance with ES standards and the ones present in the profile examples for improving performance and execution in low memory environments, for the sake of reducing problem's search space.

\subsection {Selection of Feature Combinations}

We implement the last step of \momit~by representing solutions of the miniaturization problem as bit-vectors $\vec{x}=\left\{x_1,\ldots,x_{\left| OF \right|} \in {0,1}\right\}$, where $x_j$ indicates if feature $f_j \in OF$ is included in the combination of features in the solution with $\vec{x}:x_j=1$ if it is included, $0$ otherwise.
Let $\vec{x}$ be a candidate solution and $Sel$ a function mapping a bit-vector $\vec{x}$ into the corresponding set of features $F'$, we define $UDR$ (user's dissatisfaction rate, $-USR$), $CS$ (code size), $MU$ (memory usage), and $ET$ (Execution time) as:
\begin{align*}
	UDR (\vec{x}) & =-USR (Sel(\vec{x})\cup ComF)\\
	CS (\vec{x}) & =CS(Sel(\vec{x})\cup ComF)\\
	MU (\vec{x}) & =MU(Sel(\vec{x})\cup ComF)\\
	ET (\vec{x}) & =ET(Sel(\vec{x})\cup ComF)\\
\end{align*}
\noindent where $CS (\vec{x})$, $MU (\vec{x})$ and $ET (\vec{x})$ are the code size of the JS interpreter, the memory usage, and the execution time of the JS application with the features in $\vec{x}$ and the compulsory features $ComF$. The problem to solve is to find a set of solutions $\vec{x}$ whose elements are Pareto-optimal.


\subsection{Dependencies Among JS Interpreter Features}

One of the contributions of this work is the identification of 10 hidden dependencies among the features explored.  In~\Cref{table:dependencies}, we present the arbitrary ID numbers that we assign to each  feature studied in this work,  and a brief description. These dependencies, when not considered, prevent developers and automated approaches to generate valid JS-interpreters. They come  from our preliminary study in which we evaluated each feature independently and from the evaluation of \momit, for which we run the search process with the full list of features and detected compilation errors and--or execution errors for the \textit{test harness}.

We found that the most restrictive  set of features are those related to ROM built-ins (IDs 7-10), in the sense that when they are activated, modifying the default value of other features will break the code. This is corroborated with the fact that the authors of \duktape{} commented out the lines of code that activate these features in the low-memory-environments profile to avoid errors, and provide an additional use-ROMs profile that only activate these features. Hence, we consider the dependency of ROM-built-ins partially documented.  

While some dependencies between features, which are not documented, may seem related due to their name (\eg~JSON built-in depends on JSON support, IDs 26, 27), others' dependencies are not that obvious  (\eg~\textit{Global} built-in depends on the \textit{Number} built-in, IDs 24,30).

\begin{table}[ht]
\caption{Dependencies found between the features studied in duktape}
\label{table:dependencies}
\centering
\begin{tabular}{p{0.1\columnwidth}p{0.8\columnwidth}}
\hline
Feature IDs & Description                                                                                                                                               \\ \hline
26,27       & Disabling JSON built-in requires to disable JSON support too, but not the other way around                                                                \\
32,34       & Disabling regular expression support requires to disable the string built-in as well                                                                      \\
7-10        & Storing String, and Objects in ROM requires to enable ROM Global inherit and disabling Hstring Array index feature                                        \\
11,14       & Disabling reference counting (garbage collection) requires to disable  the use of double linked heap.                                                     \\
72-74    & Minimum, Maximum and shrink limit for duktape heap string table  have to be set to the same value to avoid resizing during execution time \\
16,20       & These options provide support for augmenting ES error objects to comply with ES~5, and have to be deactivated together                                     \\
17,21       & These options Augment an ES error object at throw time and have to be deactivated together                                                                \\
31,24       & Disabling duktape object built-in requires to disable global built-in too                                                                 \\ 
24,30       & Disabling global built-in requires to disable number built-in too \\ \hline                                                                                        
\end{tabular}
\end{table}

To overcome the problem of evaluating unfeasible solutions, due to dependencies between features, we provide a mechanism to repair solutions when evaluating candidates.  The mechanism consists of detecting if any of the features with dependencies has been changed its default value;  if that is the case, we proceed to adjust the feature's values according to~\Cref{table:dependencies}.  For example, if we detect that feature 26 has been deactivated, we proceed to deactivate feature 27 as well.  These apply for all dependencies described in ~\Cref{table:dependencies}, except features 7-10 and 72-74.

When ROM built-ins features are activated, the rest of the features studied have to remain with their default values.  At least this is what we observe when testing \momit~using NSGA-II and \hrs~algorithms.  We presume that this is why the authors of \duktape~provide a separate configuration file with only these features activated, called \textit{roms-builtins.yaml}.  Hence, if we find a solution that activates one of the ROM options we either: (1) activate all ROM built-ins and reset all features values to the default JS interpreter ones, or deactivate ROM built-ins to avoid conflicts.

Finally, for feature IDs 72-74, which define the maximum, minimum and shrink limit for \duktape~heap string table, if any of these features' values are changed, from the default value, we reset them simultaneously to the minimum value suggested in the \textit{low\_memory.yaml} configuration file provided by \duktape.

We acknowledge that  the list of dependencies shown in Table~\ref{table:dependencies} is not exhaustive.  However,  it is comprehensive enough to execute the 23 JS tests selected for evaluate \momit without errors, discussed in the next Section.

\section{Evaluation of \momit}
\label{sec:evaluation}
In this section, we evaluate the effectiveness of \momit~at miniaturizing JS interpreters.  The \textit{quality focus} is the reduction of memory usage and code size to fit a JS application on more constrained devices, while keeping execution time, and user's preferences with respect to device rank in sound levels.  The \textit{perspective} is that of companies and individuals interested in porting their exiting JS code base to more constrained devices.  The \textit{context} consists of a subset of 23 JS scripts belonging to a JS benchmark  (SunSpider 1.0.2) aimed to test the core JS language only~\cite{sunspider}.  We select this testbed based on the following criteria: the JS tests do not make use of printing functions (which are not available by default on \duktape{}), third-party libraries, or any stub to execute them;  it includes a balanced and comprehensive use of JS language, including math, string processing, timing, etc.; the authors claim that they reflect real problems developers face by using JS.

In \Cref{table:sunspider} we present the tests included in our experiments, with their compulsory features, memory usage, and execution time measurements when compiling using default \duktape{} JS interpreter features. In the last column, we show the number of devices where the JS tests can be ported, based on the values from~\Cref{table:devices_preferences}.  We omit code size as this value is the same for all the JS tests when using the default features, that is  \emph{570 KB}.

\begin{table}[t]
\centering
\renewcommand{\arraystretch}{1.0}
\renewcommand{\tabcolsep}{.5mm}
\caption{JS tests used in this work taken from the Sunspider benchmark 1.0.2}
\label{table:sunspider}
\scriptsize
\begin{tabular}{llrrr}
\hline
JS test                  & $ComF$                      & $MU$ (KB)            & $ET$ (Sec.)          & Devices                    \\ \hline
3d-cube                  & 15; 29; 34;31               & 166.496              & 0.205                & 3                              \\
3d-morph                 & 15; 29                      & 132                  & 0.46                 & 3                              \\
3d-raytrace              & 15; 19; 29;31               & 387.936              & 0.25                 & 3                              \\
access-binary-trees      & 29                          & 179.44               & 0.235                & 3                              \\
access-fannkuch          & 15;31;34                    & 132.704              & 0.47                 & 3                              \\
access-nbody             & 15; 29                      & 146.848              & 0.415                & 3                              \\
access-nsieve            & 15                          & 131.296              & 0.14                 & 3                              \\
bitops-3bit-bits-in-byte &                             & 128.176              & 0.44                 & 3                              \\
bitops-bits-in-byte      &                             & 128.528              & 0.465                & 3                              \\
bitops-bitwise-and       &                             & 126.144              & 1.275                & 4                              \\
bitops-nsieve-bits       & 15                          & 130.624              & 0.775                & 3                              \\
controlflow-recursive    &                             & 184.832              & 0.21                 & 3                              \\
crypto-aes               & 15; 19; 29; 34; 32          & 178.216              & 0.21                 & 3                              \\
crypto-md5               & 15; 34;32                   & 176.544              & 0.355                & 3                              \\
crypto-sha1              & 15; 34;32                   & 163.088              & 0.34                 & 3                              \\
date-format-tofte        & 15; 19;24;31;34;32;7;8;9;10 & 1817.2               & 0.535                & 2                              \\
date-format-xparb        & 19;24;31;34;32;7;8;9;10     & 183.824              & 0.335                & 3                              \\
math-cordic              & 19                          & 132.72               & 0.475                & 3                              \\
math-partial-sums        & 29                          & 129.968              & 0.495                & 3                              \\
math-spectral-norm       & 29                          & 144.24               & 0.19                 & 3                              \\
string-base64            & 29; 34;32                   & 275.632              & 0.725                & 3                              \\
string-fasta             & 15; 34;32                   & 140.064              & 0.875                & 3                              \\
string-validate-input    & 15; 29; 34;32               & 614.464              & 1.16                 & 2                              \\ \hline
\textbf{Median}          &                             & \multicolumn{1}{l}{} & \multicolumn{1}{l}{} & \multicolumn{1}{r}{\textbf{3}} \\ \hline
\end{tabular}\end{table}

We discarded 3 files from the benchmark, which code size exceeds 100 KB.  The reason is that our approach stores the JS file in memory to execute it, and files over 100 KB already exceeds the memory capacity of one device in our list (\textit{JN5168} with 32 KB) making it impossible to find any solution that fits on it.

To perform the search we instantiate \momit~using NSGA-II algorithm with a repair function, and random search with repair function, which we refer to it as \hrs. 

\subsection{Research Questions}
To evaluate \momit, we set our case study around the following research questions (RQ)

\begin{myenumi}
	\item[\bf{(RQ1)}] \RQone \\
	The aim of this question is to quantify how much is the improvement on performance metrics after miniaturizing the JS interpreter, and based on this results to determine to how many new devices the JS code can be ported.  We compute the percentage change of the performance metrics before and after miniaturizing, using Equation~\ref{eq:delta_m} for $CS$, $MU$, and $ET$ metrics.  Next, we define $NDA$ as the difference between number of devices that we can port a JS application after and before miniaturizing.  We consider a JS application to be ported to a device $l$, if the code size and the memory usage is less or equal than the storage and memory capacity of $l$ for one or more solutions in the Pareto front.
    \item[\bf{(RQ2)}] \RQtwo \\
    The rationale behind this question is to identify which is the best algorithm to instantiate \momit~in terms of execution time, and quality of the solutions generated. To measure quality of results generated for each algorithm we compute the hypervolume (HV)~\cite{zitzler1999multiobjective} and Pareto front Size (PFS) indicators. HV provides a measure that considers the convergence and diversity of the resulting approximation set.
Higher values of the HV metric are desirable. PFS measures the number of solutions included in the Pareto front approximation comprised of the non-dominated solutions of all the algorithms evaluated, and for all the runs.  Higher PFS indicates that an algorithm scores more of the non-dominated values. To determine the significance of the obtained results, we compute the Mann--Whitney U test at $5\%$ significance level of confidence.  Mann--Whitney U is non-parametric test, \ie~ does not take into account if the samples follow a normal distribution.
   
    \end{myenumi}

\subsection {Algorithm's Control Parameter Tuning}
It is crucial for search-space exploration to tune the control parameters of the algorithms instantiated.  The main constraint that we have is respect with the evaluation of solutions on the objective space, which requires to build a JS interpreter using \duktape~python script, which takes about 20 seconds, plus the compilation of the harness code, followed by the execution of the application 10 times to average the memory and run time values which may vary between executions. Hence, we have to keep the number of evaluations and number of individuals (for EA algorithms) relatively low to produce results in a reasonable amount of time.
Since we do not have any reference of applying NSGA-II to miniaturization of JS interpreter, we ran a grid search~\cite{bergstra2012random} for the three parameters: population size ($\mu$), crossover probability (CXPB), and mutation probability (MUTPB). Our grid search space considered the following values: \{$\mu$ = [10]\} $\times$ \{CPBX = [0.6, 0.7, 0.8, 0.9]\} $\times$ \{MUTPB = [0.1, 0.15, 0.2, 0.25]\}. We use HV as the quality indicator when applying grid search.  We found the highest HV of NSGA-II with the following control parameters: $\mu$ = 10, CPBX = 0.8, MUTPB = 0.1.
To test the performance robustness and reduce the observational error we run  our experiments 30 times, and we report median values for the search grid and in the evaluation of \momit.

We use number of evaluations as the stopping criteria. As the maximum number of evaluations increase, the algorithm obtains better quality results on average. The increase in quality is usually very fast when the maximum number of evaluation is low.  That is, the slope of the curve quality versus maximum number of evaluations is high at the very beginning of the search. But this slope tends to decrease as the search progresses.  We set the number of evaluations to 250.  The rational of choosing this value is that we observed that with this value, the slope of the curve is \emph{low enough}. In our case \emph{low enough} is when we observe that it is not possible to miniaturized more a JS interpreter to fit in the most constrained IoT device from our device's list after that number of evaluations.



\section{Results}
\label{sec:results}

In this section we answer the research questions related to \momit's~evaluation.

\subsection{RQ1: \RQone}

In~\Cref{table:minresults}, we present the results obtained from \momit~after miniaturizing the JS test files included in our testbed. Columns 2-4 show the reduction percent of code size, memory usage, and execution time. Column 5 shows the number of new devices where we can fit each JS test after the miniaturization process, and Column 6 shows the total number of devices where we can deploy each JS test. These are median values obtained from the Pareto front approximation set combining the results of NSGA-II and \hrs.  

With respect to code size, we observe that 20 out of the 23 tests were reduced by more than 20\% with math-partial-sums the test with maximum reduction of 36.58\%.  

Memory usage exhibits the highest reduction of the three metrics considered, with a median of 55.51\%, reaching a maximum reduction of 92.93\% with date-format-tofte, which is the test with the most compulsory features, thus with the smaller search space, making it easier to find higher improvements.

Execution time reduction reached a median of 35.71\% with maximum reduction of 76.34\% in \textit{bitops-bits-in-byte}.  

We contrast these improvements with the ability of \momit~to miniaturize JS applications to fit in more constrained devices ($NDA$ metric). We recall that the median number of IoT devices where we can deploy the JS tests (\Cref{table:sunspider}) is three, using \duktape~default configuration values, but now is four devices out of five.   The JS tests that were $NDA$ reached more than one device are \textit{date-format-tofte} and \textit{string-validate-input}. For these two tests, the memory usage was higher than the capacity of all of the SoC devices considered before performing the miniaturization and thanks to \momit~it was possible to port them to most of them. For \textit{string-validate-input}, beside the median memory improvement is only 11\%, there are five solutions (out of 66) on the Pareto front that improve memory usage by more than 70\%. If we found at least one solution that fits in a new device, we report it in $NDA$, and Devices columns.

On the other hand, there are three tests that, despite the reductions achieved, could not be fit in more constrained devices because it was not possible to reduce the footprint of the JS interpreter. They are \textit{3d-raytrace}, \textit{bitops-bitwise-and}, and \textit{string-base64}. For the first and third tests, the memory improvement achieved was not enough to reach the capacity of the \emph{Photon} device. For the second one, the original test already fits on four of the five considered devices. Hence, fitting \textit{bitops-bitwise-and} on the highest constrained device would require reducing by more than half the original code size and use one third of the memory used with the default values, at the same time.  
\momit~was not able to port any JS test to \textit{JN5168} micro-controller due to its low memory and storage capacity.



\hypobox{We conclude that \momit~can successfully improve performance metrics of JS interpreters by removing unnecessary features to execute JS applications in more constrained devices without modifying their original source code.}

\begin{table}[ht]
\caption{Median results of the miniaturization process using \momit~on 23 JS tests}
\label{table:minresults}
\centering
\begin{tabular}{lrrrrr}
\hline
JS test & $\delta CS$ & $\delta MU$ & $\delta ET$ & $NDA$ & Devices \\ \hline
3d-cube & -5.67 & -55.51 & -31.71 & 1 & 4 \\
3d-morph & -12.56 & -67.26 & -34.78 & 1 & 4 \\
3d-raytrace & -28.19 & -15.89 & -8 & 0 & 3 \\
access-binary-trees & -7.64 & -49.63 & -31.91 & 1 & 4 \\
access-fannkuch & -30.13 & -53.94 & -36.7 & 1 & 4 \\
access-nbody & -30.67 & -50.93 & -37.35 & 1 & 4 \\
access-nsieve & -30.94 & -68.45 & -35.71 & 1 & 4 \\
bitops-3bit-bits-in-byte & 24.72 & -82.61 & -27.27 & 1 & 4 \\
bitops-bits-in-byte & -33.62 & -69.53 & -76.34 & 1 & 4 \\
bitops-bitwise-and & -34.57 & -75.94 & -34.12 & 0 & 4 \\
bitops-nsieve-bits & -31.21 & -66.62 & -34.19 & 1 & 4 \\
controlflow-recursive & -33.89 & -63.58 & -19.05 & 1 & 4 \\
crypto-aes & -28.97 & 4.56 & -26.6 & 1 & 4 \\
crypto-md5 & -31.59 & -35.54 & -60.56 & 1 & 4 \\
crypto-sha1 & -24.58 & -51.31 & -41.18 & 1 & 4 \\
date-format-tofte & -31.03 & -92.93 & -55.14 & 2 & 4 \\
date-format-xparb & -30.59 & -25.78 & -37.31 & 1 & 4 \\
math-cordic & -35.57 & -61.37 & -43.16 & 1 & 4 \\
math-partial-sums & -36.58 & -62.58 & -49.49 & 1 & 4 \\
math-spectral-norm & -9.06 & -63.18 & -42.11 & 1 & 4 \\
string-base64 & -32.1 & -22.8 & -24.14 & 0 & 3 \\
string-fasta & -28.62 & -50.11 & -37.14 & 1 & 4 \\
string-validate-input & -29.53 & -11.34 & -12.43 & 2 & 4 \\ \hline
\textbf{Total Median} & \textbf{-30.59} & \textbf{-55.51} & \textbf{-35.71} & \textbf{1} & \textbf{4} \\ \hline
\end{tabular}
\end{table}

\subsection{RQ2: \RQtwo}


In Table~\ref{table:et250ev}, we present the time measured for the two algorithm implementations. We did not include $p-values$ of the Mann--Whitney U Tests because \textbf{all the differences are statistically significant with large Cliff's $\delta$ effect size.}

We observe that NSGA-II perform faster than \hrs. We suggest that this is the result of generating more repeated solutions that we do not need to evaluate again-and-again in comparison with \hrs~and thus explore less the search space. As in both implementations, we store in a dictionary the objective values achieved by every solution evaluated to reduce computation time.  Hence, when an algorithm finds repeated solutions, we retrieve the objective values stored in the dictionary avoiding the expensive process of recompiling a JS interpreter and executing the JS test again.  That is the the algorithm \textit{per se} is not the bottleneck in \momit~process pipeline, but the generation of \duktape~interpreter, and the compilation and execution of the JS test.

\begin{table}[ht]
\caption{Execution time of \momit's search algorithms in seconds. (All the result differences  are statistically significant with large Cliff's $\delta$ effect size.)}
\label{table:et250ev}
\centering
\begin{tabular}{lrr}
\hline
JS test                   & NSGA-II & \hrs \\ \hline
3d-cube                  & 2642    & 5035      \\ 
3d-morph                 & 2987    & 5948      \\ 
3d-raytrace              & 2679    & 4253      \\ 
access-binary-trees      & 2419    & 5004      \\ 
access-fannkuch          & 3198    & 6511      \\ 
access-nbody             & 2606    & 5375      \\ 
access-nsieve            & 3163    & 6272      \\ 
bitops-3bit-bits-in-byte & 2547    & 4800      \\ 
bitops-bits-in-byte      & 2571    & 4982      \\ 
bitops-bitwise-and       & 3462    & 6843      \\ 
bitops-nsieve-bits       & 2834    & 5619      \\ 
controlflow-recursive    & 2340    & 4778      \\ 
crypto-aes               & 2836    & 4702      \\ 
crypto-md5               & 2786    & 3978      \\ 
crypto-sha1              & 2785    & 4009      \\ 
date-format-tofte        & 2461    & 4375      \\ 
date-format-xparb        & 2536    & 3974      \\ 
math-cordic              & 2857    & 4261      \\ 
math-partial-sums        & 2879    & 4122      \\ 
math-spectral-norm       & 2534    & 3736      \\ 
string-base64            & 3947    & 6328      \\ 
string-fasta             & 3285    & 4833      \\ 
string-validate-input    & 9793    & 16791     \\ 
\hline
\textbf{Total Median}             & \textbf{2786}    & \textbf{4833}      \\ \hline
\end{tabular}
\end{table}

In \Cref{table:numberParetoSolutions}, we present the numbers of non-dominated solutions contributed by each algorithm to the Pareto Front approximation. We observe that the numbesr of solutions contributed by \hrs{} overcome those of NSGA-II, indicating bad performance of the latter one, according to the PFS metric. Finally, in \Cref{table:HVmeanValues}, we report the average hypervolume values for each algorithm. The HV metric measures the convergence and diversity of the solutions found and shows that \hrs{} overcome NSGA-II in terms of quality.

\begin{table}[ht]
\caption{Pareto optimal solutions found by NSGA-II and \hrs.}
\label{table:numberParetoSolutions}
\centering
\begin{tabular}{lccc}
\hline
Script                   & Solutions & NSGAII       & \hrs           \\
\hline
3d-cube                  & 102       & 39 (38.24\%) & 63 (61.76\%) \\
3d-morph                 & 73        & 35 (47.95\%) & 38 (52.05\%) \\
3d-raytrace              & 81        & 3 (3.70\%)   & 78 (96.30\%)      \\
access-binary-trees      & 89        & 47 (52.81\%) & 42 (47.19\%) \\
access-fannkuch          & 66        & 7 (10.61\%)  & 59 (89.39\%) \\
access-nbody             & 44        & 13 (29.55\%) & 31 (70.45\%) \\
access-nsieve            & 33        & 8 (24.24\%)  & 25 (75.76\%) \\
bitops-3bit-bits-in-byte & 71        & 29 (40.85\%) & 42 (59.15\%) \\
bitops-bits-in-byte      & 40        & 10 (25.00\%)    & 30 (75.00\%)    \\
bitops-bitwise-and       & 36        & 9 (25.00\%)     & 27 (75.00\%)    \\
bitops-nsieve-bits       & 52        & 7 (13.46\%)  & 45 (86.54\%) \\
controlflow-recursive    & 40        & 7 (17.50\%)   & 33 (82.50\%)  \\
crypto-aes               & 58        & 6 (10.34\%)   & 52 (89.66\%)      \\
crypto-md5               & 37        & 3 (8.11\%)   & 34 (91.89\%) \\
crypto-sha1              & 63        & 6 (9.52\%)   & 57 (90.48\%) \\
date-format-tofte        & 24        & 2 (8.33\%)   & 22 (91.67\%) \\
date-format-xparb        & 66        & 4 (6.06\%)   & 62 (93.94\%) \\
math-cordic              & 55        & 9 (16.36\%)  & 46 (83.64\%) \\
math-partial-sums        & 36        & 1 (2.78\%)   & 35 (97.22\%) \\
math-spectral-norm       & 53        & 1 (1.89\%)   & 52 (98.11\%) \\
string-base64            & 55        & 11 (20.00\%)    & 44 (80.00\%)    \\
string-fasta             & 37        & 1 (2.70\%)    & 36 (97.30\%)  \\
string-validate-input    & 66        & 11 (16.67\%) & 55 (83.33\%) \\
\hline
\end{tabular}
\end{table}

\begin{table}[ht]
\caption{Average HV values in 30 runs, Mann\textemdash Whitney U Test, and Cliff$'$s $\delta$ Effect Size (ES).}
\label{table:HVmeanValues}
\centering
\begin{tabular}{lcccc}
\hline
Script                   & NSGA-II & \hrs   & $p-value$ & $ES$ \\
\hline
3d-cube                  & 0.78   & 0.88 & 2.31E-10      & large  \\
3d-morph                 & 0.85   & 0.90  & 5.34E-08      & large  \\
3d-raytrace              & 0.56   & 0.83   & 1.69E-17
      & large     \\
access-binary-trees      & 0.74   & 0.80  & 3.03E-03      & medium \\
access-fannkuch          & 0.49   & 0.70  & 2.49E-13      & large  \\
access-nbody             & 0.27   & 0.37 & 8.00E-06      & large  \\
access-nsieve            & 0.33   & 0.46 & 3.02E-05      & large  \\
bitops-3bit-bits-in-byte & 0.19   & 0.38 & 3.43E-08      & large  \\
bitops-bits-in-byte      & 0.19   & 0.34 & 1.09E-06      & large  \\
bitops-bitwise-and       & 0.28   & 0.44 & 5.18E-09      & large  \\
bitops-nsieve-bits       & 0.45   & 0.61 & 1.09E-06      & large  \\
controlflow-recursive    & 0.17   & 0.32 & 7.24E-10      & large  \\
crypto-aes               & 0.77   & 0.91   & 1.69E-17
      & large     \\
crypto-md5               & 0.42   & 0.74 & 1.69E-17      & large  \\
crypto-sha1              & 0.43   & 0.72 & 2.03E-16      & large  \\
date-format-tofte        & 0.34   & 0.78 & 1.69E-17      & large  \\
date-format-xparb        & 0.34   & 0.80  & 1.69E-17      & large  \\
math-cordic              & 0.23   & 0.44 & 9.75E-14      & large  \\
math-partial-sums        & 0.21   & 0.46 & 1.69E-17      & large  \\
math-spectral-norm       & 0.17   & 0.38 & 3.21E-16      & large  \\
string-base64            & 0.87   & 0.92 & 1.02E-07      & large  \\
string-fasta             & 0.37   & 0.84 & 1.69E-17      & large  \\
string-validate-input    & 0.84   & 0.92 & 1.09E-09      & large \\
\hline
\end{tabular}
\end{table}

We also compare the $NDA$ metric for NSGA-II and \hrs{} to see whether the differences in quality enable \hrs{} to port the JS interpreter to more devices and observer that only in two tests, \textit{date-format-tofte} and \textit{string-validate-input}, does \hrs{} overcome NSGA-II.

These results show that \hrs~overcome NSGA-II, which is a state-of-the-art evolutionary algorithms. One reason could be the limited number of evaluations and the small population size, which might not allow NSGA-II to perform to its best. To verify whether the low performance of NSGA-II compared to \hrs{} is the result of few evaluations and considering the high cost incurred when augmenting the numbers of evaluations, we decided to run a microexperiment using only two randomly-selected JS tests from our testbed and applying both NSGA-II and \hrs~ with the following control parameters: population 100 (NSGA-II), 15,000 evaluations for both algorithms, with 10 independent runs (due to the excessive computation time). To simplify this experiment, we evaluate only the first three performance objectives (Code size, Memory usage and execution time). We include an additional search algorithm to our microexperiment SWAY~\cite{8249828} because its authors claim that it can be useful in situations where evaluating solutions in the search space is expensive but there is a high correlation between the decision space and the search space. 

SWAY performs most of the evaluations in the decision space and limits the number of objective evaluations compared to EAs. SWAY samples from a large set of feasible solutions and cluster them based on the value of their decision variables, to perform a reduced amount of objective evaluations instead of generating an initial population of random solutions and evolving the most prominent ones during a determined number of generations.

To instantiate SWAY using a binary representation, we formulate the problem of miniaturization as a software product line (SPL), where each combination of features leads to the creation of a new product (\ie~a JS interpreter). The population is generated after defining a list of Boolean predicates that models the dependency between JS interpreter features and the compulsory features for each selected JS test, using the conjunctive normal form (CNF). The predicates define what is a valid solution. The next step is to input the Boolean predicates into a SAT solver, which is responsible to find set of all satisfiable CNF expressions, limited to the size of the population. Then, SWAY clusters the population according to their decision variables using a radial coordinate system, \ie{} clustering the individuals by the numbers of ones or zeros contained in their binary string representations. For more details, we refer the readers to the original source~\cite{8249828}.

We instantiate SWAY as a blackbox, by clonning the original implementation of SWAY from the author's version control system~\footnote{https://github.com/ginfung/FSSE}. We wrote a class to represent the problem of miniaturization and another to define the experiment settings. The control parameters for SWAY are a population of 15,000 (5,000  individuals more than in the original publication) and 10 independent runs like NSGA-II and \hrs~respectively.

In \Cref{table:micro-experiment_metrics_new}, we present the execution time and quality metrics of the solutions found in the microexperiment. We compare the results of executing 250 evaluations against 15,000, except for SWAY.  The reason is that SWAY requires larger populations to be effective, and in the previous experiments we limit the population of NSGA-II to 10 individuals.  Additionally, we recomputed HV metric considering only 3 objectives to make a fair comparison between the 250 and the 15,000 evaluations.

With respect to execution time, SWAY is the fastest approach, taking only 8 and 11 minutes of (median values), which is expected as SWAY performs less objective evaluations compared to the other algorithms. It is followed by NSGA-II with execution time around 27 and 42 hours. Then, \hrs~follows with 67 and 105 hours approximately. For the HV metric, the results are very similar than for 250 evaluations, where \hrs~outperformed NSGA-II, while SWAY is below both of them.  Note that HV metric for NSGA-II considerably improved from 0.49 to 0.82 in \textit{access-fannkuch} JS test,  while in \textit{3d-cube} remained the same.  Concerning PFS metric, again \hrs{}~reports the higher gain of non-dominated solutions added to the Pareto Front; NSGA-II contribute few dominated solutions to the Pareto Front and SWAY did not contribute any. Based on these results, we suggest that increasing the number of evaluations for NSGA-II and \hrs{} was not fruitful, but impractical from the point of view of practitioners, especially because the number of new devices reached after miniaturizing remained the same (Column 5), that when performing 250 evaluations.  Yet, there is an improvement in HV and PFS metrics by increasing the number of evaluations for NSGA-II and \hrs.  Although SWAY performed faster than the rest of the algorithms, it could not reach the same amount of new devices for \textit{3d-cube}; it did not contribute any new solutions to the Pareto Front, which means that the solutions generated by SWAY were all dominated by the other approaches. 


\begin{table*}[ht]
\caption{Execution time and quality metrics of the microexperiment. ET and HV are median and mean values, respectively. 
}
\label{table:micro-experiment_metrics_new}
\centering
\begin{tabular}{llccccccc}
\hline
\multirow{2}{*}{Script}          & \multirow{2}{*}{EA} & \multicolumn{3}{c}{250 evaluations} & \multicolumn{3}{c}{15,000 evaluations} & \multirow{2}{*}{NDA} \\
                                 &                     & ET (sec.)     & HV.mean    & PFS    & ET (sec.)     & HV.mean     & PFS      &                      \\
\hline
\multirow{3}{*}{3d-cube}         & NSGA-II              & 2,642              & 0.67       & 4      &        96,321       & 0.67        & 7        & 1                    \\
                                 & \hrs                  &  5,035             & 0.90       & 55     &   240,132            & 0.94        & 107      & 1                    \\
                                 & SWAY                & NA            & NA         & NA     & 488           & 0.49        & 0.00     & 0                    \\
\hline
\multirow{3}{*}{access-fannkuch} & NSGA-II              &   3,198            & 0.49       & 2      &     149,514          & 0.82        & 15       & 1                    \\
                                 & \hrs                  &      6,511         & 0.73       & 51     &    377,487           & 0.85        & 69       & 1                    \\
                                 & SWAY                & NA            & NA         & NA     & 680           & 0.24        & 0.00     & 1 
\\
\hline
\end{tabular}
\end{table*}

After considerably augmenting the number of evaluations and obtaining similar results, we suggest that the inferior performance of NSGA-II in comparison with \hrs{} is caused by the transformation operators (crossover and mutation) employed, which are typically used for binary solutions, but not necessary the most suitable ones for this particular problem.  For example, we could have a mutation operator that considers dependencies among features, and mutate them simultaneously to produce a valid JS interpreter. However, we are not sure to what extent this could improve the results obtained.  Hence, we left the definition and evaluation of new transformation operators for the miniaturization of JS interpreters for future work.

With respect to SWAY, we are surprised by the low quality of the obtained results. We suggest this is the result of assuming that there is a direct relationship between the decision variables and the objective values, \ie{} the fact that SWAY clusters solutions based on the numbers of ones/zeros might be appropriate for SPLs where "one" means adding a feature/component and "zero" otherwise.  On the other hand, in the context of \duktape~features the concept of "one" equal to adding extra functionality does not hold.  For example, consider feature 5 from \Cref{table:prelim_major_concern}, which default value is  "zero", and when switch it to "one" removes ES compliance to reduce memory usage.  In other words, having more "ones" does not imply adding more functionality to the interpreter in all cases.

\hypobox{We conclude that the best algorithm to instantiate \momit~is \hrs{} due to the quality of the solutions obtained and the number of devices it enabled us to port a JS interpreter. NSGA-II performs faster than \hrs{} and can reach almost the same number of devices than \hrs. Finally, if execution time is a main concern, SWAY is the best algorithm although its solutions are of lower quality.}

\section{Discussion}
\label{sec:momit-discussion}

In this section, we discuss and put in perspective the results obtained in the experiments after evaluating \momit. We also take a step back and look at the overall approach and its implementation.

During the evaluation of \momit, the authors of this work observed that some programming practices can make certain JS interpreter features mandatory, whose presence can prevent greater improvements in performance. For example, in \textit{date-format-tofte} script, the use of JS \textit{Date prototype} property conflicts with the use of ROM built-ins. \textit{Date prototype} allows adding new properties and methods to the \textit{Date()} object. However, allowing such addition conflicts with the ROM feature, which makes this property read-only when activated. Hence, \duktape~configuration script would fail to produce the source and headers of the new JS interpreter for \textit{date-format-tofte} with the ROM built-ins enabled. To avoid generating invalid solutions, we set ROM built-ins as a compulsory feature using JS interpreter default values.

The benefits of applying \momit{} can be extended to other programming languages by providing the PRs of the applications to be executed. For example CPython, a C implementation of Python, has been manually customized to generate Python interpreters for embedded devices (\eg~\textit{PyMite, TinyPython}, etc.) by removing unessential features and supporting just a subset of Python syntax. Using \momit{}, developers do not need to spend the time and effort in miniaturizing CPython but could focus on their applications. The designs and implementations of programming languages tend to favour interpreters/virtual machines and hence would be amenable to \momit{}.

Miniaturization, in general principle, is concerned about reducing the storage, memory, and CPU requirements of applications. Therefore, previous approaches, like MoMS~\cite{ali2011moms} and others, focused on miniaturizing the applications themselves, without considering their run-time support. Therefore, they work well for applications compiled directly to run on embedded systems. They would perform less well on interpreted/virtual-machine applications because they do not consider this support. In addition, MoMS (and such approaches) and \momit{} are complementary it two ways. First, we could apply MoMS on the applications themselves while we could apply \momit{} to identify, in an interpreter, pre-requirements pertaining to performance, their mapping to the interpreter implementation, and then \momit{} to choose the most appropriate combination of these features.

We are aware that it is possible to reduce code size by applying processor-specific compiler options. For example, we compared the compilation of \textit{harness} using default GCC options, as we used in our experiments (\Cref{lst:gcc-default-harness}), against adding optimized options for ARM-processor devices (\Cref{lst:gcc-flags-harness}) and observed that code size went down from 362 KB to 132 KB. The choice of the compilation options would complement \momit{}. We considered the ``default'' options because we prefer to leave the tuning of these options to developers to keep our implementation as general as possible without losing its benefits and generality.

\begin{lstlisting}[,language=bash, caption=Compiling \textit{harness} using GCC with default options options,label={lst:gcc-default-harness},escapeinside={(*@}{@*)},basicstyle=\scriptsize,numbers=none,xleftmargin=15pt,breaklines=true]
gcc -std=c99 -o harness harness.c duktape.c -lm
\end{lstlisting}

\begin{lstlisting}[,language=bash, caption=Compiling \textit{harness} using GCC with optimized flags for reducing code size on ARM devices,label={lst:gcc-flags-harness},escapeinside={(*@}{@*)},basicstyle=\scriptsize,numbers=none,xleftmargin=15pt,breaklines=true]
gcc -o harness -m32 -std=c99 -Wall -Os  -fomit-frame-pointer -flto  -fno-asynchronous-unwind-tables  -ffunction-sections -Wl,--gc-sections 	-fno-stack-protector -Iduktape-src duktape.c harness.c -lm
\end{lstlisting}

IoT devices come in many shapes and forms. In recent years, some IoT devices have been steadily promoted to full-fledged computers thank to advances in computing hardware and battery technologies. In particular, the \textit{RPI} is being used in many hobbyist, industrial, and research IoT projects. Hence, developers could consider that they have available large storage and memory as well as powerful CPUs for their IoT project. However, \textit{RPI}s and other such computers remain computers that can hardly be used in many scenarios because of their form factors, their energy consumptions, their costs, and environment considerations. For example, it is not environmentally-friendly, cost-effective, and technically feasible to drop hundreds or thousands of \textit{RPI}s in forests to monitor droughts and fires. Consequently, \textit{RPI}s would remain at the ``edge'' of an IoT application monitoring forests and developers would use \momit{} to miniaturize their applications to run on smaller, cheaper, cleaner devices.

\section{Threats to validity}
\label{sec:threats-validity}

This section discusses the threats to validity of our study following common guidelines for empirical studies~\cite{Yin}.

\emph{Construct validity threats} concern the relation between theory and observation. They can be due to imprecision in the measurements performed in the study.  We measured code size, memory usage and execution using Linux's well-known commands, and repeat the measurements 10 times for each single JS application execution, additionally to the 30 executions when applying \momit~in~\Cref{sec:evaluation}. 
As in most previous studies we cannot exclude the impact of the operating system. What is measured is a mix of Linux and application
actions. We mitigate this by running the application multiple times, and executing the experiments in a dedicated \textit{RPI} machine, disconnected from Internet, and just with the necessary tools and scripts used for the experiments.

In the preliminary study as well as in the evaluation of~\momit, the JS interpreter is compiled and the JS applications executed during the experiments, so we can assure that the resultant JS interpreters are valid.

\emph{Threats to internal validity} concern our selection of JS testbed, tools and analysis method.  In this study we used a particular yet representative subset of JS test files belonging to a benchmark for JS applications as a proxy for JS applications.  Regarding performance metrics measurement we use well know theory and measurements were repeated several times to ensure statistical validity.

\emph{Conclusion validity threats} concern the relation between the treatment and the outcome. We paid attention not to violate assumptions of the constructed statistical models. In particular, we used a non-parametric test, Mann-Whitney U Test, Cliff's $\delta$ ES, that does not make assumptions on the underlying data distribution.

\emph{Reliability validity threats} concern the possibility of replicating this study. The applications tools used in this study are open-source and can be access with the data collected an generated in the online replication package~\cite{MoMItRep} .

\emph{Threats to external validity} concern the possibility to generalize our results. These results have to be interpreted carefully as they may depend on on the specific device where we
ran the experiments, the operating system, the JS applications used, their compulsory features, and the level of compliance to ES that the company is interested to maintain, which we keep it to the minum to improve performance as much as possible.  Although, we use the example of miniaturization of JS interpreters, other interpreters (Python, Lua, etc.) could be miniaturized using the process described by \momit~by executing a PR analysis and with help of an in interpreter that can be customized to (de)activate optional features, like CPython.


\section{Conclusion and Future Work}
\label{sec:conclusion}

In this work, we presented \momit, an automated multi-objective approach for miniaturizing JS applications to run on constrained IoT devices. This approach supports IoT companies interested in deploying their applications on different devices, while accounting for based on typical performance metrics (storage and memory usage, and CPU time) and developers' preferences (\eg{} prices, with or without network interfaces).

First, we performed a preliminary study to identify, from a set of \textbf{283} JS interpreter features including ECMA script and interpreter-specific ones, those features that have an impact on the performance of the resulting executable JS interpreter in terms of storage usage, memory usage, and CPU time. We thus identified \textbf{86} features.  

Second, by considering the 86 identified features in the previous step, we formulated a multi-objective approach, \momit, and implement it using NSGA-II and hybrid-random-search and SWAY algorithms to miniaturize a JS interpreter based on the compulsory features of any JS application. 

Finally, we evaluated \momit~by miniaturizing 23 JS tests from a well-known JS testbed (SunSpider). We showed that \momit~can reduce code size, memory usage, and CPU time by median values of 31\%, 56\%, and 36\% respectively. Moreover, it managed to miniaturize 21 JS test applications out of 23 to the size of a quarter coin (\photon~and \textit{ESP32}).

While developing and evaluting \momit, we also identified 10 hidden dependencies between 20 of the 86 features studied, which were not documented, and the correction of a software bug~\footnote{https://github.com/svaarala/duktape/issues/2002} that affected the compilation of the JS interpreter.

We released the source code of \momit~as open-source, so researchers and practitioners can benefit of our work and replicate of our results~\cite{MoMItRep}.


The steps followed by \momit~are generally enough to be applied for miniaturizing any code interpreter, providing the list PRs of the application to be executed, and the list of customizable interpreter features.

As a future work, we plan to evaluate \momit~on other code interpreters like Python, or Java. Another interesting direction of research is the detection of dependencies between features automatically. 

\ifCLASSOPTIONcompsoc
\section*{Acknowledgments}
\else
\section*{Acknowledgment}
\fi
This work has been supported by the Natural Sciences and Engineering Research Council of Canada (NSERC).

\ifCLASSOPTIONcaptionsoff
  \newpage
\fi

\balance
\bibliographystyle{IEEEtran}

\begin{IEEEbiography}[{\includegraphics[width=1in,height=1.25in,clip,keepaspectratio]{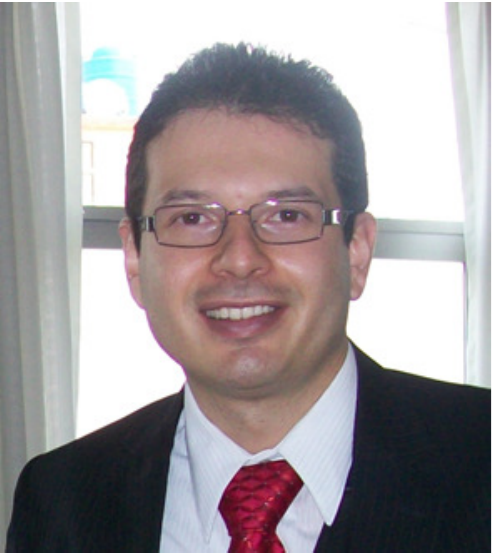}}]{Rodrigo Morales} is a Postdoctoral fellow at Concordia University working on new methods to improve the quality of software development for the Internet of Things.  He obtained his BS. degree in computer science in 2005 from Polytechnic of Mexico. In 2008, he obtained his MS. in computer technology from the same University, where he also worked as a Professor in the computer Science department for five years. He has also worked in the bank industry as a software developer for more than three years.  He obtained his Ph.D. degree in computer engineering from Polytechnic of Montreal where he earned the best thesis award of 2017.  He has published in top software engineering Journals and like IEEE TSE, ESEM, and JSS and top conferences including ICSE, and SANER.  He is one of the main organizers of the 1st International Workshop on Software Engineering Research \& Practices for the Internet of Things (SERP4IoT), co-located with ICSE 2019, and actively participate as committee member of ICSME and ICPC conferences.

His research interests include software development for the internet of things, software design quality, energy efficiency, automated-refactoring, anti-patterns, and mobile apps.
\end{IEEEbiography}

\begin{IEEEbiography}[{\includegraphics[width=1in,height=1.25in,clip,keepaspectratio]{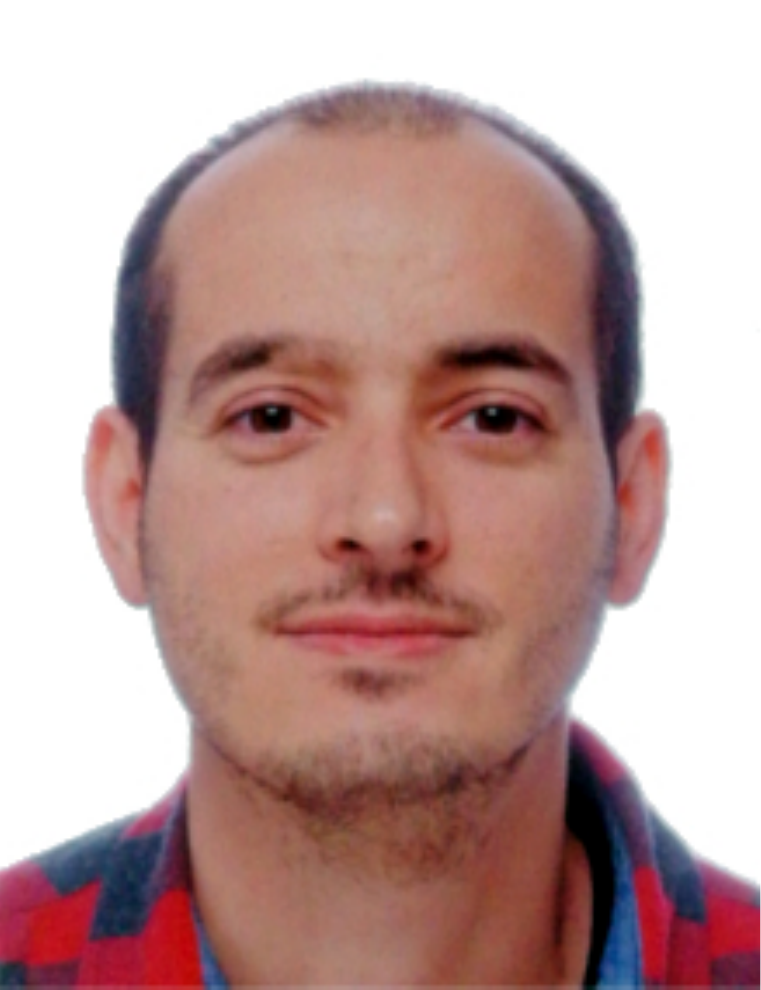}}]{Rub\'{e}n Saborido} is a researcher at the Networking and Emerging Optimization group at the Department of Computer Science at University of Malaga (Spain). He received his BS. degree in Computer Engineering and his MS. in Software Engineering and Artificial Intelligence from University of Malaga (Spain), where he worked for three years as a researcher assistant. In 2017, he received a Ph.D. in Computer Engineering from Polytechnique Montréal (Canada) and his thesis was nominated for best thesis award. In 2018 he held a postdoctoral fellowship at Concordia University (Canada), where he worked on search-based software engineering for the Internet of Things (IoT) Rub\'{e}n research focuses on search-based software engineering for IoT. He is also interested in the use of metaheuristics to solve multidisciplinary real-world problems of interest for our society and computer science. He has published several papers in ISI indexed journals (such as EMSE, IEEE TSE, and Evolutionary Computation) and conference papers in IEEE ICPC, MCDM, IEEE SANER, and ACM ESEC/FSE. He has co-organized the International Conference on Multiple Criteria Decision Making, in 2013. He is on the organizing committee of the 1st International Workshop on Software Engineering Research \& Practices for the Internet of Things (SERP4IoT), co-located with ICSE 2019. He is also on the application committee of the Real World Applications (RWA) track of the Genetic and Evolutionary Computation Conference (GECCO), from 2016 up today.
\end{IEEEbiography}

\begin{IEEEbiography}[{\includegraphics[width=1in,height=1.25in,clip,keepaspectratio]{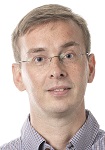}}]{Yann-Ga{\"e}l Gu{\'e}h{\'e}neuc} is full professor at the Department of Computer Science and Software Engineering of Concordia University since 2017, where he leads the Ptidej team on evaluating and enhancing the quality of the software systems, focusing on the Internet of Things and researching new theories, methods, and tools to understand, evaluate, and improve the development, release, testing, and security of such systems. Prior, he was faculty member at Polytechnique Montréal and Université de Montréal, where he started as assistant professor in 2003. In 2014, he was awarded the NSERC Research Chair Tier II on Patterns in Mixed-language Systems. In 2013-2014, he visited KAIST, Yonsei U., and Seoul National University, in Korea, as well as the National Institute of Informatics, in Japan, during his sabbatical year. In 2010, he became IEEE Senior Member. In 2009, he obtained the NSERC Research Chair Tier II on Software Patterns and Patterns of Software. In 2003, he received a Ph.D. in Software Engineering from University of Nantes, France, under Professor Pierre Cointe's supervision. His Ph.D. thesis was funded by Object Technology International, Inc. (now IBM Ottawa Labs.), where he worked in 1999 and 2000. In 1998, he graduated as engineer from École des Mines of Nantes. His research interests are program understanding and program quality, in particular through the use and the identification of recurring patterns. He was the first to use explanation-based constraint programming in the context of software engineering to identify occurrences of patterns. He is interested also in empirical software engineering; he uses eye-trackers to understand and to develop theories about program comprehension. He has published papers in international conferences and journals, including IEEE TSE, Springer EMSE, ACM/IEEE ICSE, IEEE ICSME, and IEEE SANER. He was the program co-chair and general chair of several events, including IEEE SANER'15, APSEC'14, and IEEE ICSM'13.
\end{IEEEbiography}

\begin{table*}[ht]
\centering
\renewcommand{\arraystretch}{1.0}
\renewcommand{\tabcolsep}{.5mm}
\caption{List of \duktape~features used in this work. }
\label{table:features}
\scriptsize
\begin{tabular}{llrrlr}
\hline
ID & Property & Default value & Modified value & Duktape Category & Bin. value \\ \hline
1 & DUK\_USE\_ALLOW\_UNDEFINED\_BEHAVIOR & FALSE & TRUE & Platform and portability options & 0 \\
2 & DUK\_USE\_FATAL\_MAXLEN & 128 & 64 & Platform and portability options & 1 \\
3 & DUK\_USE\_EXEC\_PREFER\_SIZE & FALSE & TRUE & low Memory management options & 0 \\
4 & DUK\_USE\_LEXER\_SLIDING\_WINDOW & TRUE & FALSE & low Memory management options & 1 \\
5 & DUK\_USE\_LIGHTFUNC\_BUILTINS & FALSE & TRUE & low Memory management options & 0 \\
6 & DUK\_USE\_PREFER\_SIZE & FALSE & TRUE & low Memory management options & 0 \\
7 & DUK\_USE\_ROM\_STRINGS & FALSE & TRUE & low Memory management options & 0 \\
8 & DUK\_USE\_ROM\_OBJECTS & FALSE & TRUE & low Memory management options & 0 \\
9 & DUK\_USE\_ROM\_GLOBAL\_INHERIT & FALSE & TRUE & low Memory management options & 0 \\
10 & DUK\_USE\_HSTRING\_ARRIDX & TRUE & FALSE & low Memory management options & 1 \\
11 & DUK\_USE\_REFERENCE\_COUNTING & TRUE & FALSE & Garbage collection options & 1 \\
12 & DUK\_USE\_PARANOID\_ERRORS & FALSE & TRUE & ECMAScript Edition 5 (ES5) options & 0 \\
13 & DUK\_USE\_FUNC\_NAME\_PROPERTY & TRUE & FALSE & ECMAScript Edition 5 (ES5) options & 1 \\
14 & DUK\_USE\_DOUBLE\_LINKED\_HEAP & TRUE & FALSE & ECMAScript Edition 5 (ES5) options & 1 \\
15 & DUK\_USE\_ARRAY\_BUILTIN & TRUE & FALSE & ECMAScript Edition 5 (ES5) options & 1 \\
16 & DUK\_USE\_AUGMENT\_ERROR\_CREATE & TRUE & FALSE & ECMAScript Edition 5 (ES5) options & 1 \\
17 & DUK\_USE\_AUGMENT\_ERROR\_THROW & TRUE & FALSE & ECMAScript Edition 5 (ES5) options & 1 \\
18 & DUK\_USE\_BOOLEAN\_BUILTIN & TRUE & FALSE & ECMAScript Edition 5 (ES5) options & 1 \\
19 & DUK\_USE\_DATE\_BUILTIN & TRUE & FALSE & ECMAScript Edition 5 (ES5) options & 1 \\
20 & DUK\_USE\_ERRCREATE & TRUE & FALSE & ECMAScript Edition 5 (ES5) options & 1 \\
21 & DUK\_USE\_ERRTHROW & TRUE & FALSE & ECMAScript Edition 5 (ES5) options & 1 \\
22 & DUK\_USE\_FUNCTION\_BUILTIN & TRUE & FALSE & ECMAScript Edition 5 (ES5) options & 1 \\
23 & DUK\_USE\_FUNC\_FILENAME\_PROPERTY & TRUE & FALSE & ECMAScript Edition 5 (ES5) options & 1 \\
24 & DUK\_USE\_GLOBAL\_BUILTIN & TRUE & FALSE & ECMAScript Edition 5 (ES5) options & 1 \\
25 & DUK\_USE\_JC & TRUE & FALSE & ECMAScript Edition 5 (ES5) options & 1 \\
26 & DUK\_USE\_JSON\_BUILTIN & TRUE & FALSE & ECMAScript Edition 5 (ES5) options & 1 \\
27 & DUK\_USE\_JSON\_SUPPORT & TRUE & FALSE & ECMAScript Edition 5 (ES5) options & 1 \\
28 & DUK\_USE\_JX & TRUE & FALSE & ECMAScript Edition 5 (ES5) options & 1 \\
29 & DUK\_USE\_MATH\_BUILTIN & TRUE & FALSE & ECMAScript Edition 5 (ES5) options & 1 \\
30 & DUK\_USE\_NUMBER\_BUILTIN & TRUE & FALSE & ECMAScript Edition 5 (ES5) options & 1 \\
31 & DUK\_USE\_OBJECT\_BUILTIN & TRUE & FALSE & ECMAScript Edition 5 (ES5) options & 1 \\
32 & DUK\_USE\_REGEXP\_SUPPORT & TRUE & FALSE & ECMAScript Edition 5 (ES5) options & 1 \\
33 & DUK\_USE\_SOURCE\_NONBMP & TRUE & FALSE & ECMAScript Edition 5 (ES5) options & 1 \\
34 & DUK\_USE\_STRING\_BUILTIN & TRUE & FALSE & ECMAScript Edition 5 (ES5) options & 1 \\
35 & DUK\_USE\_TRACEBACKS & TRUE & FALSE & ECMAScript Edition 5 (ES5) options & 1 \\
36 & DUK\_USE\_VERBOSE\_ERRORS & TRUE & FALSE & ECMAScript Edition 5 (ES5) options & 1 \\
37 & DUK\_USE\_PC2LINE & TRUE & FALSE & ECMAScript Edition 5 (ES5) options & 1 \\
38 & DUK\_USE\_VERBOSE\_EXECUTOR\_ERRORS & TRUE & FALSE & ECMAScript Edition 5 (ES5) options & 1 \\
39 & DUK\_USE\_BYTECODE\_DUMP\_SUPPORT & TRUE & FALSE & API options & 1 \\
40 & DUK\_USE\_BASE64\_SUPPORT & TRUE & FALSE & Codecs & 1 \\
41 & DUK\_USE\_HEX\_SUPPORT & TRUE & FALSE & Codecs & 1 \\
42 & DUK\_USE\_DUKTAPE\_BUILTIN & TRUE & FALSE & Duktape specific options & 1 \\
43 & DUK\_USE\_BUFFEROBJECT\_SUPPORT & TRUE & FALSE & ECMAScript 2015 (ES6) options & 1 \\
44 & DUK\_USE\_ES6 & TRUE & FALSE & ECMAScript 2015 (ES6) options & 1 \\
45 & DUK\_USE\_ES6\_PROXY & TRUE & FALSE & ECMAScript 2015 (ES6) options & 1 \\
46 & DUK\_USE\_ES6\_UNICODE\_ESCAPE & TRUE & FALSE & ECMAScript 2015 (ES6) options & 1 \\
47 & DUK\_USE\_HTML\_COMMENTS & TRUE & FALSE & ECMAScript 2015 (ES6) options & 1 \\
48 & DUK\_USE\_SHEBANG\_COMMENTS & TRUE & FALSE & ECMAScript 2015 (ES6) options & 1 \\
49 & DUK\_USE\_REFLECT\_BUILTIN & TRUE & FALSE & ECMAScript 2015 (ES6) options & 1 \\
50 & DUK\_USE\_SYMBOL\_BUILTIN & TRUE & FALSE & ECMAScript 2015 (ES6) options & 1 \\
51 & DUK\_USE\_ES7 & TRUE & FALSE & ECMAScript 2016 (ES7) options & 1 \\
52 & DUK\_USE\_ES7\_EXP\_OPERATOR & TRUE & FALSE & ECMAScript 2016 (ES7) options & 1 \\
53 & DUK\_USE\_ES8 & TRUE & FALSE & ECMAScript 2017 (ES8) options & 1 \\
54 & DUK\_USE\_ES9 & TRUE & FALSE & ECMAScript 2018 (ES9) options & 1 \\
55 & DUK\_USE\_ENCODING\_BUILTINS & TRUE & FALSE & ECMAScript 2018 (ES9) options & 1 \\
56 & DUK\_USE\_ARRAY\_FASTPATH & TRUE & FALSE & Perfomance options & 1 \\
57 & DUK\_USE\_ARRAY\_PROP\_FASTPATH & TRUE & FALSE & Perfomance options & 1 \\
58 & DUK\_USE\_BASE64\_FASTPATH & TRUE & FALSE & Perfomance options & 1 \\
59 & DUK\_USE\_CACHE\_ACTIVATION & TRUE & FALSE & Perfomance options & 1 \\
60 & DUK\_USE\_CACHE\_CATCHER & TRUE & FALSE & Perfomance options & 1 \\
61 & DUK\_USE\_FAST\_REFCOUNT\_DEFAULT & TRUE & FALSE & Perfomance options & 1 \\
62 & DUK\_USE\_HEX\_FASTPATH & TRUE & FALSE & Perfomance options & 1 \\
63 & DUK\_USE\_HOBJECT\_HASH\_PROP\_LIMIT & 8 & 64 & Perfomance options & 1 \\
64 & DUK\_USE\_HSTRING\_LAZY\_CLEN & TRUE & FALSE & Perfomance options & 1 \\
65 & DUK\_USE\_IDCHAR\_FASTPATH & TRUE & FALSE & Perfomance options & 1 \\
66 & DUK\_USE\_JSON\_QUOTESTRING\_FASTPATH & TRUE & FALSE & Perfomance options & 1 \\
67 & DUK\_USE\_JSON\_DECSTRING\_FASTPATH & TRUE & FALSE & Perfomance options & 1 \\
68 & DUK\_USE\_JSON\_DECNUMBER\_FASTPATH & TRUE & FALSE & Perfomance options & 1 \\
69 & DUK\_USE\_JSON\_EATWHITE\_FASTPATH & TRUE & FALSE & Perfomance options & 1 \\
70 & DUK\_USE\_LITCACHE\_SIZE & 256 & FALSE & Perfomance options & 1 \\
71 & DUK\_USE\_REGEXP\_CANON\_BITMAP & TRUE & FALSE & Perfomance options & 1 \\
72 & DUK\_USE\_STRTAB\_MINSIZE & 1024 & 128 & Perfomance options & 1 \\
73 & DUK\_USE\_STRTAB\_MAXSIZE & 268435456 & 128 & Perfomance options & 1 \\
74 & DUK\_USE\_STRTAB\_SHRINK\_LIMIT & 6 & 0 & Perfomance options & 1 \\
75 & DUK\_USE\_STRTAB\_GROW\_LIMIT & 17 & 65536 & Perfomance options & 1 \\
76 & DUK\_USE\_VALSTACK\_GROW\_SHIFT & 2 & FALSE & Perfomance options & 1 \\
77 & DUK\_USE\_VALSTACK\_SHRINK\_CHECK\_SHIFT & 2 & FALSE & Perfomance options & 1 \\
78 & DUK\_USE\_VALSTACK\_SHRINK\_SLACK\_SHIFT & 4 & FALSE & Perfomance options & 1 \\
79 & DUK\_USE\_VALSTACK\_UNSAFE & FALSE & TRUE & Perfomance options & 0 \\
80 & DUK\_USE\_DEBUG\_BUFSIZE & 65536 & 2048 & Debugger options & 1 \\
81 & DUK\_USE\_COROUTINE\_SUPPORT & TRUE & FALSE & Execution options & 1 \\
82 & DUK\_USE\_PERFORMANCE\_BUILTIN & TRUE & FALSE & Performance API (High Resolution Time) & 1 \\
83 & DUK\_USE\_VOLUNTARY\_GC & TRUE & FALSE & Garbage collection options & 1 \\
84 & DUK\_USE\_FASTINT & FALSE & TRUE & Performance options & 0 \\
85 & DUK\_USE\_JSON\_STRINGIFY\_FASTPATH & FALSE & TRUE & Performance options & 0 \\
86 & DUK\_USE\_REGEXP\_CANON\_WORKAROUND & FALSE & TRUE & Performance options & 0 \\ \hline\end{tabular}
\end{table*}

\begin{table*}[ht]
\centering
\renewcommand{\arraystretch}{1.0}
\renewcommand{\tabcolsep}{.5mm}
\caption{Results of preliminary study of JS interpreter features and their impact on performance metrics. }
\label{table:prelim-complete}
\begin{tabular}{lrrrrrrr}
\hline
id & value & \textit{harness} size & mem. us. & $\delta CS$ & $\delta MU$ & median $ET$ & median $\delta ET$ \\ \hline
1 & TRUE & 555896 & 104816 & 0 & 0 & 0.71 & -13.41 \\
2 & 64 & 555896 & 104816 & 0 & 0 & 0.7 & -14.63 \\
3 & TRUE & 490824 & 104816 & -11.71 & 0 & 0.895 & 9.15 \\
4 & FALSE & 555888 & 104816 & 0 & 0 & 0.7 & -14.63 \\
5 & TRUE & 555896 & 65584 & 0 & -37.43 & 0.715 & -12.805 \\
6 & TRUE & 551696 & 104784 & -0.76 & -0.03 & 1.595 & 94.51 \\
11 & FALSE & 518176 & 115952 & -6.79 & 10.62 & 0.72 & -12.2 \\
12 & TRUE & 555976 & 104816 & 0.01 & 0 & 0.705 & -14.02 \\
13 & FALSE & 555896 & 103440 & 0 & -1.31 & 0.71 & -13.41 \\
15 & FALSE & 546424 & 99728 & -1.7 & -4.85 & 0.69 & -15.85 \\
16 & FALSE & 555728 & 104816 & -0.03 & 0 & 0.705 & -14.02 \\
17 & FALSE & 555840 & 104816 & -0.01 & 0 & 0.7 & -14.63 \\
18 & FALSE & 555776 & 104080 & -0.02 & -0.7 & 0.705 & -14.02 \\
19 & FALSE & 555280 & 93152 & -0.11 & -11.13 & 0.71 & -13.41 \\
20 & FALSE & 555896 & 104816 & 0 & 0 & 0.715 & -12.805 \\
21 & FALSE & 555896 & 104816 & 0 & 0 & 0.71 & -13.41 \\
22 & FALSE & 555728 & 103616 & -0.03 & -1.14 & 0.72 & -12.2 \\
23 & FALSE & 555896 & 104720 & 0 & -0.09 & 0.71 & -13.41 \\
24 & FALSE & 555320 & 102192 & -0.1 & -2.5 & 0.775 & -5.49 \\
25 & FALSE & 555896 & 104816 & 0 & 0 & 0.755 & -7.93 \\
26 & FALSE & 555792 & 104160 & -0.02 & -0.63 & 0.83 & 1.22 \\
28 & FALSE & 555680 & 104816 & -0.04 & 0 & 0.79 & -3.66 \\
29 & FALSE & 549800 & 97696 & -1.1 & -6.79 & 0.71 & -13.41 \\
30 & FALSE & 555432 & 102640 & -0.08 & -2.08 & 0.705 & -14.02 \\
31 & FALSE & 555144 & 99120 & -0.14 & -5.43 & 0.705 & -14.02 \\
33 & FALSE & 555904 & 104816 & 0 & 0 & 0.71 & -13.41 \\
34 & FALSE & 546280 & 98384 & -1.73 & -6.14 & 0.72 & -12.2 \\
35 & FALSE & 555840 & 104816 & -0.01 & 0 & 0.705 & -14.02 \\
36 & FALSE & 543328 & 104816 & -2.26 & 0 & 0.7 & -14.63 \\
37 & FALSE & 555736 & 104640 & -0.03 & -0.17 & 0.7 & -14.63 \\
38 & FALSE & 555896 & 104816 & 0 & 0 & 0.71 & -13.41 \\
39 & FALSE & 547216 & 104816 & -1.56 & 0 & 0.7 & -14.63 \\
40 & FALSE & 555896 & 104816 & 0 & 0 & 0.71 & -13.41 \\
41 & FALSE & 555896 & 104816 & 0 & 0 & 0.71 & -13.41 \\
42 & FALSE & 555536 & 101568 & -0.06 & -3.1 & 0.71 & -13.41 \\
43 & FALSE & 532664 & 82400 & -4.18 & -21.39 & 0.705 & -14.02 \\
44 & FALSE & 550952 & 100720 & -0.89 & -3.91 & 0.715 & -12.805 \\
45 & FALSE & 551488 & 104448 & -0.79 & -0.35 & 0.705 & -14.02 \\
46 & FALSE & 555896 & 104816 & 0 & 0 & 0.705 & -14.02 \\
47 & FALSE & 555896 & 104816 & 0 & 0 & 0.71 & -13.41 \\
48 & FALSE & 555896 & 104816 & 0 & 0 & 0.715 & -12.805 \\
49 & FALSE & 555680 & 102192 & -0.04 & -2.5 & 0.705 & -14.02 \\
50 & FALSE & 555896 & 104816 & 0 & 0 & 0.71 & -13.41 \\
51 & FALSE & 555896 & 104816 & 0 & 0 & 0.71 & -13.41 \\
52 & FALSE & 551800 & 104816 & -0.74 & 0 & 0.715 & -12.805 \\
53 & FALSE & 555768 & 103808 & -0.02 & -0.96 & 0.71 & -13.415 \\
54 & FALSE & 555896 & 104816 & 0 & 0 & 0.71 & -13.41 \\
55 & FALSE & 555424 & 102768 & -0.08 & -1.95 & 0.72 & -12.2 \\
56 & FALSE & 555744 & 104816 & -0.03 & 0 & 0.71 & -13.41 \\
57 & FALSE & 555760 & 104816 & -0.02 & 0 & 0.735 & -10.37 \\
58 & FALSE & 555808 & 104816 & -0.02 & 0 & 0.71 & -13.41 \\
59 & FALSE & 555832 & 104592 & -0.01 & -0.21 & 0.71 & -13.41 \\
60 & FALSE & 555840 & 104704 & -0.01 & -0.11 & 0.72 & -12.2 \\
61 & FALSE & 535880 & 104816 & -3.6 & 0 & 0.705 & -14.02 \\
62 & FALSE & 555808 & 104816 & -0.02 & 0 & 0.7 & -14.63 \\
63 & 64 & 555896 & 100832 & 0 & -3.8 & 0.705 & -14.02 \\
64 & FALSE & 555880 & 104816 & 0 & 0 & 0.71 & -13.41 \\
65 & FALSE & 555848 & 104816 & -0.01 & 0 & 0.7 & -14.63 \\
66 & FALSE & 555888 & 104816 & 0 & 0 & 0.71 & -13.41 \\
67 & FALSE & 555848 & 104816 & -0.01 & 0 & 0.71 & -13.41 \\
68 & FALSE & 555840 & 104816 & -0.01 & 0 & 0.695 & -15.24 \\
69 & FALSE & 555840 & 104816 & -0.01 & 0 & 0.71 & -13.41 \\
70 & FALSE & 555896 & 104816 & 0 & 0 & 0.7 & -14.63 \\
71 & FALSE & 555840 & 104816 & -0.01 & 0 & 0.71 & -13.41 \\
74 & 0 & 555896 & 104816 & 0 & 0 & 0.705 & -14.02 \\
75 & 65536 & 555896 & 104816 & 0 & 0 & 0.71 & -13.41 \\
76 & FALSE & 555896 & 104384 & 0 & -0.41 & 0.71 & -13.41 \\
77 & FALSE & 555896 & 104352 & 0 & -0.44 & 0.7 & -14.63 \\
78 & FALSE & 555896 & 104816 & 0 & 0 & 0.7 & -14.63 \\
79 & TRUE & 555896 & 104816 & 0 & 0 & 0.7 & -14.63 \\
80 & 2048 & 555896 & 104816 & 0 & 0 & 0.71 & -13.41 \\
81 & FALSE & 551568 & 103872 & -0.78 & -0.9 & 0.7 & -14.63 \\
82 & FALSE & 555848 & 104304 & -0.01 & -0.49 & 0.705 & -14.02 \\
83 & FALSE & 555848 & 104816 & -0.01 & 0 & 0.725 & -11.59 \\
84 & TRUE & 592960 & 104816 & 6.67 & 0 & 0.63 & -23.17 \\
85 & TRUE & 560144 & 104816 & 0.76 & 0 & 0.715 & -12.805 \\
86 & TRUE & 687016 & 104816 & 23.59 & 0 & 0.7 & -14.63 \\\hline
\end{tabular}
\end{table*}

\begin{table*}[ht]
\centering
\renewcommand{\arraystretch}{1.0}
\renewcommand{\tabcolsep}{.5mm}
\caption{Results of preliminary study of JS interpreter features and their impact on performance metrics (features with dependencies). }
\label{table:prelim-complete-dependencies}
\begin{tabular}{lrrrrrrr}
\hline
id & value & \textit{harness} size & mem. us. & $\delta CS$ & $\delta MU$ & median $ET$ & median $\delta ET$ \\ \hline
11\_14 & FALSE & 514080 & 115632 & -7.52 & 10.32 & 0.72 & -12.2 \\
26\_27 & FALSE & 537264 & 104160 & -3.35 & -0.63 & 0.71 & -13.41 \\
32\_34 & FALSE & 519208 & 96656 & -6.6 & -7.79 & 0.72 & -12.2 \\
7 to 10 & vary & 696048 & 12656 & 25.21 & -87.93 & 0.71 & -13.41 \\
72\_73 & 128 & 555736 & 97648 & -0.03 & -6.84 & 0.705 & -14.02 \\ \hline
\end{tabular}
\end{table*}

\end{document}